# Diversity-Multiplexing Tradeoff in Multi-Antenna Multi-Relay Networks: Improvements and Some Optimality Results

Shahab Oveis Gharan, Alireza Bayesteh, and Amir K. Khandani

Coding & Signal Transmission Laboratory
Department of Electrical & Computer Engineering
University of Waterloo
Waterloo, ON, N2L 3G1
shahab, alireza, khandani@cst.uwaterloo.ca

## Abstract

This paper investigates the benefits of Amplify-and-Forward (AF) relaying in the setup of multi-antenna wireless networks. Reference [1] introduced the idea of *Random Sequential* (RS) relaying and showed that it achieves the maximum diversity gain in a general multi-antenna network. Here, we show that random unitary matrix multiplication at the relay nodes empowers the RS scheme to achieve a better Diversity-Multiplexing Tradeoff (DMT) as compared to the traditional AF relaying. First, we study the case of a multi-antenna full-duplex single-relay two-hop network, for which we show that the RS achieves the optimum DMT. Applying this result, we derive a new achievable DMT for the case of multi-antenna half-duplex parallel relay network. Interestingly, it turns out that the DMT of the RS scheme is optimum for the case of multi-antenna two parallel non-interfering half-duplex relays. Next, we show that random unitary matrix multiplication also improves the DMT of the Non-Orthogonal AF relaying scheme of [2] in the case of a multi-antenna single relay channel. Finally, we study the general case of multi-antenna full-duplex relay networks and derive a new lower-bound on its DMT using the RS scheme[1].

## I. Introduction

Recently, cooperative schemes have been proposed as candidates to exploit the spatial diversity offered by the relay networks (see for example [2], [5]–[9] and references therein). *Decode-and-Forward* (DF), *Amplify-and-Forward* (AF) and *Compress-and-Forward* (CF) are the primary relaying strategies discussed in the literature. While DF and CF strategies are popular for small-scale networks to achieve or approach capacity values (see for example [10]–[12] and references therein), the AF relaying turns out to be more suitable in realizing cooperative transmission strategies (see for example [1], [2], [5], [8], [13] and references therein).

Financial supports provided by Nortel, and the corresponding matching funds by the Federal government: Natural Sciences and Engineering Research Council of Canada (NSERC) and Province of Ontario (ORF-RE) are gratefully acknowledged.

[1]The materials of this paper are (in part) reported in *Library and Archives Canada Technical Report* [3], July 2008, and in the *46th Allerton Conference on Communication, Control, and Computing* [4], September 2008.

In AF relaying, the relays do not decode the transmitted messages. Instead, they forward their received signals to the destination after a proper amplification. Hence, the relays require less computing power and the end-to-end system experiences a smaller delay in comparison with the other relaying strategies. Accordingly, AF relaying schemes are suitable for deployment in practical wireless systems. Moreover, in contrast to DF relaying, the performance of AF relaying is not limited by the source-to-relay channel quality. Also, unlike CF relaying, parallel relays can be used in conjunction with the AF strategy to realize the power boosting offered by coherently combining the received signals at the destination [9]. Indeed, it is shown that AF relaying is asymptotically (large number of relays) optimal (capacity achieving) for both single-antenna [14] and multi-antenna [9], [15] parallel relay networks.

While AF relaying is investigated well in single-antenna networks, much less is known about its application in multi-antenna networks. Indeed, unlike the single-antenna scenario, in this case the AF multipliers are matrices rather than scalars. Hence, finding the optimum AF matrices becomes challenging.

A fundamental measure to evaluate the performance of cooperative schemes is the DMT, which was first introduced by Zheng and Tse in the context of point-to-point multi-antenna fading channels [16]. Roughly speaking, the diversity-multiplexing tradeoff identifies the optimal compromise between the *transmission reliability* and the *data rate* in the high Signal to Noise Ratio (SNR) regime.

The Non-orthogonal Amplify-and-Forward (NAF) scheme, first proposed by Nabar *et al.* in [17], has been further studied by Azarian *et al.* in [2] for the single-antenna multi-relay setup. In addition to analyzing the DMT of the NAF scheme, reference [2] shows that NAF is the best in the class of AF strategies for single-antenna single-relay systems. However, the NAF scheme falls far from the upper-bound in multi-relay systems.

Yang and Belfiore in [13] study the DMT performance of the NAF scheme for the multi-antenna parallel relay setup. Moreover, based on the non-vanishing determinant criterion, the authors constructed a family of space-time codes for the NAF scheme over multi-antenna channels. However, as shown in [13], the NAF scheme falls far from the DMT upper-bound in the multiple-antenna setup, particularly for small values of multiplexing gain. Indeed, even for the case of a multi-antenna two-hop single-relay setup, the NAF scheme is unable to achieve the maximum diversity gain of the system.

Recently, [1] has studied the achievable DMT of the AF relaying for a general multi-relay network. For this purpose, they propose a new AF relaying scheme which is called RS. They show that the RS scheme achieves the optimum diversity gain in a general multi-antenna multi-relay network. However, they established the DMT optimality results only for the single-antenna multi-relay network. Also, the authors in [18] independently obtained the same DMT optimality result using a different approach.

Yuksel *et al.* in [6] apply CF strategy and show that CF achieves the DMT upper-bound for multi-antenna half-duplex single-relay networks. However, in their proposed scheme, the relay node needs to know the Channel State Information (CSI) of all the channels in the network, which may not be practical. More recently, Avestimehr *et al.* in [12] show that a variant of the CF relaying achieves the capacity of any general single-antenna Gaussian relay network within a fixed number of bits, which only depends on the number of nodes in the network. Furthermore, the authors in [19] show that the result of [12] is still valid for both multi-antenna Gaussian relay networks with or without fast (ergodic) Rayleigh fading.

In this paper, we investigate the benefits of AF relaying in multi-antenna multi-relay networks. For this purpose, we study the application of the RS scheme proposed in [1]. The key elements of the proposed scheme are: 1) signal transmission through sequential paths in the network, 2) path timing such that no non-causal interference is caused from the transmitted signal of the future paths on the received signal of the current path, 3) multiplication by a random unitary matrix at each relay node, and 4) no signal boosting in AF relaying at the relay nodes, i.e. the received signal is amplified by a coefficient with the absolute value of at most 1. We derive the DMT of the RS scheme for multi-antenna multi-relay networks. To accomplish this task, we first study a simple structure, namely the multi-antenna full-duplex two-hop single-relay network. We show that unlike the traditional AF relaying, the RS scheme achieves the optimum DMT. This fact can be justified as follows: using the traditional AF relaying, there exists a chance that the eigenvectors corresponding to the largest eigenvalues of the incoming channel matrix of the relay project to the eigenvectors corresponding to the small eigenvalues of the relay's outgoing channel matrix. This event degrades the performance of traditional AF relaying in the multi-antenna setup. However, in the RS scheme, due to the random independent unitary matrix multiplication at the relay nodes for different time-slots, such an event is much less likely to happen. This fact will be elaborated throughout the paper.

Next, we study the case of multi-antenna half-duplex parallel relay network and, by deriving its DMT, we show that the RS scheme improves the DMT of the traditional AF relaying scheme. Interestingly, it turns out that the DMT of the RS scheme is optimum for the multi-antenna half-duplex parallel two-relay ($K = 2$) setup with no direct link between the relays. We also show that utilizing random unitary matrix multiplication improves the DMT of the NAF relaying scheme of [2] in the case of a multi-antenna single relay channel.

Finally, we study the class of general full-duplex multi-antenna relay networks whose underlying graph is *directed acyclic* and all nodes are equipped with the same number of antennas. Using the RS scheme, we derive a new lower-bound for the achievable DMT of this class of networks. It turns out that the new DMT lower-bound meets the optimum DMT at the corner points, corresponding to the maximum multiplexing gain and the maximum diversity gain of the network, respectively. Another point worth mentioning is that the RS scheme is *robust* in the sense that it achieves all points of the DMT curve with no modification of the underlying parameters. In other words, the relay nodes of the network perform the same operation, no matter at which point of the DMT curve the scheme is operating.

The rest of the paper is organized as follows. System model is introduced in section II. Section III is dedicated to Diversity-Multiplexing Tradeoff analysis of the RS scheme in the multi-antenna setup. This section is further divided into four subsections as follows. Subsection III-A studies the multi-antenna single-relay two-hop network and also the multi-antenna multi-hop relay network with one relay in each hop. Subsection III-B is dedicated to the multi-antenna half-duplex parallel relay network. The well-known multi-antenna single-relay channel (with direct link between the source and the destination) is investigated in subsection III-C. Subsection III-D studies the achievable DMT of the RS scheme for the general multi-antenna full-duplex relay networks whose underlying graph is directed acyclic. Finally, section IV concludes the paper.

*A. Notations*

Throughout the paper, the superscripts $^T$ and $^H$ stand for matrix operations of transposition and conjugate transposition, respectively. Capital bold letters represent matrices, while lowercase bold letters and regular letters represent vectors and scalars, respectively. $\|\mathbf{v}\|$ denotes the norm of vector $\mathbf{v}$, while $\|\mathbf{A}\|$ represents the Frobenius norm of matrix $\mathbf{A}$. $|\mathbf{A}|$ denotes the determinant of matrix $\mathbf{A}$ and $\mathrm{Tr}\{\mathbf{A}\}$ denotes the trace of matrix $\mathbf{A}$. $\log(.)$ denotes the base-2 logarithm. The notation $\mathbf{A} \preccurlyeq \mathbf{B}$ is equivalent to $\mathbf{B} - \mathbf{A}$ being a positive semi-definite matrix. Motivated by the definition in [16], we define the notation $f(P) \doteq g(P)$ as $\lim_{P\to\infty} \frac{f(P)}{\log(P)} = \lim_{P\to\infty} \frac{g(P)}{\log(P)}$. Similarly, $f(P)\dot{\leq}g(P)$ and $f(P)\dot{\geq}g(P)$ are equivalent to $\lim_{P\to\infty} \frac{f(P)}{\log(P)} \leq \lim_{P\to\infty} \frac{g(P)}{\log(P)}$ and $\lim_{P\to\infty} \frac{f(P)}{\log(P)} \geq \lim_{P\to\infty} \frac{g(P)}{\log(P)}$, respectively. Finally, we use $A \approx B$ to denote the approximate equality between $A$ and $B$, such that by substituting $A$ with $B$ the validity of the equations are not compromised.

## II. SYSTEM MODEL

In this paper, we study the DMT for various setups of multiple-antenna relay networks. The common assumptions for all setups are:

- The network consists of a single source, one or multiple relays and a single destination

- All nodes in the network are assumed be equipped with multiple antennas. However, the number of antennas at each node are denoted differently depending on the scenario being discussed. In the setup of Theorem 1 in subsection III-A, which studies the DMT of a two-hop single-relay network, the number of antennas at the source, relay and destination are denoted by $m$, $p$, and $n$, respectively. However, in Theorem 2 of this section which considers the setup of multi-hop relay network, the number of antennas at nodes are denoted by $N_0, N_1, \cdots, N_h$ where $h$ denotes the total number of hops. In subsection III-B, in which the setup of a parallel relay network is considered, the number of antennas at the source and the destination are assumed to be $m$ and $n$. In Theorem 3, $K$ parallel relays with the same number (denoted by $p$) of antennas are considered while Theorem 4 studies the setup of two parallel relays with different number of antennas denoted by $n_1$ and $n_2$ while assuming $n = m$. The general setup of a single-relay network (with direct link between the source and the destination) is studies in subsection III-C (Theorem 5) and the number of antennas at the source, relay and destination are denoted by $m$, $p$, and $n$, respectively. Finally, the general setup of a multi-antenna full-duplex relay network with directed acyclic underlying graph is considered in subsection III-D (Theorem 6) in which the number of antennas at node $i$ is denoted by $N_i$.

- All channels in the network are assumed to be circularly symmetric zero-mean unit-variance Gaussian with gains that remain unchanged during the entire transmission period (quasi-static Rayleigh fading). For each link, the receiver side is assumed to have perfect knowledge about the corresponding channel and the transmitter side is assumed to have no knowledge about the channel.

- The noise vector at each node (relay or destination) is assumed to be additive white with circularly symmetric zero-mean unit-variance Gaussian distribution.

- The relays are assumed to work in full-duplex mode in sub-sections III-A and III-D and half-duplex mode in sub-sections

III-B and III-C.

- The source and the relay(s) have power constraints $\mathbb{E}\{\mathbf{x}_t^H \mathbf{x}_t\} \leq P$ and $\mathbb{E}\{\mathbf{x}_r^H \mathbf{x}_r\} \leq P$, respectively, where the average is taken with respect to $\mathbf{x}_t$ for the source and with respect to $\mathbf{x}_r$ and $\mathbf{y}_r$ (the received signal at the relay) for the relays[2]. As the main focus of this paper is the analysis of DMT, the source and the relay(s) are assumed to operate in the high SNR regime ($P \rightarrow \infty$).

## III. Diversity-Multiplexing Tradeoff

### A. Two-Hop Single Relay Network

This setup corresponds to the network consisting of a source, a single full-duplex relay and a destination with no direct link between the source and the destination. The source, relay, and destination are equipped with $m$, $p$, and $n$ antennas, respectively (see Figure 1). The channel between the source and the relay is denoted by $\mathbf{H}$ and the channel between the relay and the receiver is denoted by $\mathbf{G}$. First, we study the achievable DMT by the traditional AF scheme in Lemmas 1 and 2 and show the achievable DMT in general does not match with the optimal value, which is achievable by the DF scheme[3]. Then, in Theorem 1 we prove that using the proposed RS scheme, which is a modification of the traditional AF scheme, the optimal DMT is indeed achievable. Theorem 2 generalizes the result of Theorem 1 to the case of multi-hop relay network and shows that the proposed RS scheme still achieves the optimum DMT provided that a certain relationship between the number of antennas at nodes is satisfied.

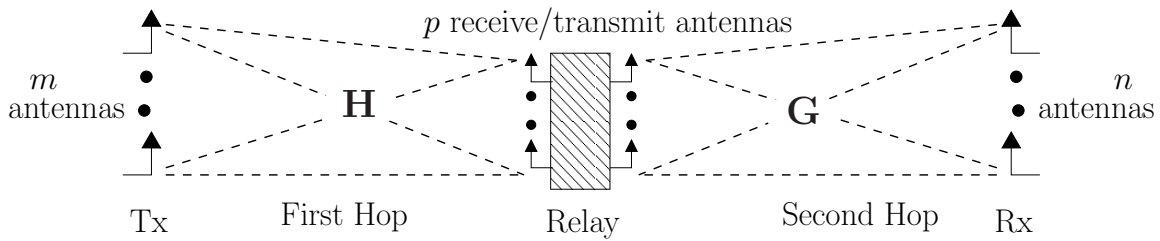

Fig. 1. Schematic of a multi-antenna single-relay two-hop network

In the traditional AF strategy, the received signal at the relay is multiplied by a constant $\alpha$ such that the power constraint at the relay is satisfied and then it is transmitted to the destination. The corresponding received signal at the destination can be written as

$$\mathbf{y} = \alpha \mathbf{G}\mathbf{H}\mathbf{x}_t + \alpha \mathbf{G}\mathbf{n}_r + \mathbf{n}_d, \tag{1}$$

where $\mathbf{x}_t$ denotes the transmitted signal from the source and $\mathbf{n}_r \sim \mathcal{CN}(\mathbf{0}, \mathbf{I}_p)$ and $\mathbf{n}_d \sim \mathcal{CN}(\mathbf{0}, \mathbf{I}_n)$ denote the noise vectors at the relay and at the destination, respectively.

---

[2]Note that as the channels are assumed to be fixed during the whole transmission period, the expectation are not taken with respect to any of the channel matrices.

[3]In fact, this configuration is a special case of the degraded relay channel studied by [11]. In [11], the authors show that the DF scheme achieves the capacity of the degraded relay channel.

**Lemma 1** *The DMT of the system given in (1) is upper-bounded by the DMT of the following system:*

$$\mathbf{y} = \alpha \mathbf{GH} \mathbf{x}_t + \mathbf{n}_d, \tag{2}$$

*and is lower-bounded by the DMT of the following system:*

$$\mathbf{y} = \alpha \mathbf{GH} \mathbf{x}_t + \sqrt{c \log(P) + 1} \mathbf{n}_d, \tag{3}$$

*for some constant c.*

*Proof:* See Appendix I. ∎

**Lemma 2** *The DMT of the systems in (2) and (3) are equal.*

*Proof:* See Appendix II. ∎

A direct conclusion of the Lemmas 1 and 2 is that the DMT of the two-hop network can be expressed as the DMT of the product channel $\mathbf{GH}$ which is computed in [13]. Due to the result given in Proposition 1 in [13], assuming $m, n \geq p$, the DMT of the product channel $\mathbf{A} = \mathbf{GH}$ is a piecewise-linear function connecting the points $(r, d_{\mathbf{A}}(r)), r = 0, 1, \ldots, p$, where

$$d_{\mathbf{A}}(r) = (p - r)(q - r) - \frac{1}{2} \left\lfloor \frac{[(p - \Delta - r)^+]^2}{2} \right\rfloor, \tag{4}$$

$q = \min(m, n)$ and $\Delta = |m - n|$. On the other hand, the piecewise-linear function connecting the integer points $(r, (p - r)(q - r))$ can be easily derived as the upper-bound by considering each of the source-relay or the relay-destination cuts. Comparing (4) with the upper-bound, it follows that the traditional AF scheme achieves the optimum DMT only when $r \geq p - \Delta$. This motivates us to use a variant of AF scheme which achieves the optimum DMT in all cases. In fact, using the traditional AF scheme, there are three sources of outage: (i) the outage in the source-relay link, (ii) the outage in the relay-destination link, and (iii) the projection of the eigenmodes of $\mathbf{H}$ over the eigenmodes of $\mathbf{G}$ is very small. More precisely, the matrix $\mathbf{V}^H(\mathbf{G})\mathbf{U}(\mathbf{H})$, in which $\mathbf{V}^H(\mathbf{G})$ denotes the right eigenvector matrix from the Singular Value Decomposition (SVD) of $\mathbf{G}$ and $\mathbf{U}(\mathbf{H})$ denotes the left eigenvector matrix from the SVD of $\mathbf{H}$, has very small eigenvalues. The extra term $\frac{1}{2} \left\lfloor \frac{[(p - \Delta - r)^+]^2}{2} \right\rfloor$ in (4) is due to the third source of outage. The first two outage events depend on the distribution of the eigenvalues of $\mathbf{H}$ and $\mathbf{G}$, while the third event depends solely on the *direction* of the eigenvectors of these two matrices. This suggests that in order to eliminate the extra terms in $\frac{1}{2} \left\lfloor \frac{[(p - \Delta - r)^+]^2}{2} \right\rfloor$ one can multiply the received signal at the relay by $\alpha \mathbf{\Theta}$, for some $p \times p$ unitary matrix $\mathbf{\Theta}$ (for preserving the power constraint at the relay). However, it should be noted that if $\mathbf{\Theta}$ does not change across the transmission block, the performance of the systems does not change. Therefore, we propose that in each transmission slot an independent unitary matrix $\mathbf{\Theta}_l$ is used and at the destination side the decoding is performed across $L$ transmission slots[4]. This is exactly what is being done in the proposed RS scheme in [1]. Indeed, this setup is a simple example of the general setup of the relay network studied in [1] in which the source and the destination are connected through a single path. In this case the proposed RS scheme reduces to the following: the source's message is sent using $L$ slots through the same path; at the relay side the received signal is multiplied by a randomly independent (through different slots) unitary matrix, and subsequently, it is multiplied by a scalar $\alpha \leq 1$[5] such that the power constraint is satisfied, and the result is transmitted in the next slot. At the destination, following receiving the signal of the slots $2, 3, \ldots, L + 1$, the source message is decoded. In the following theorem, we show that as long as $L$ is above a certain threshold, the probability of the third outage event is negligible compared to the first two outage events and hence, the optimum DMT is achievable by the RS scheme.

---

[4]From practical point of view, the transmission slots are assumed to be long enough to make the probability of error solely dominated by the outage event. This fact is more elaborated in [16].

**Theorem 1** *Consider a two-hop network consisting of a source with $m$ antennas and a destination with $n$ antennas that are connected through a full-duplex relay with $p$ antennas. Let us define $q = \min(m, n)$. Providing $L$ is large enough such that $L \geq \min^2(p, q) \max(p, q)$, the RS scheme achieves the optimum DMT, which is the piecewise-linear function connecting the points $(k, (p - k)(q - k)), k = 0, 1, \ldots, \min(p, q)$.*

*Proof:* Using Lemmas 1 and 2, the DMT of the system using the proposed RS scheme is equal to the DMT of the following system:

$$\mathbf{Y} = \alpha \mathbf{\Omega} \mathbf{X}_t + \mathbf{N}_d, \tag{5}$$

where $\mathbf{X}_t \triangleq [\mathbf{x}_t(1), \cdots, \mathbf{x}_t(L)]^T$, $\mathbf{Y} = [\mathbf{y}(1), \cdots, \mathbf{y}(L)]^T$, and $\mathbf{N}_d = [\mathbf{n}_d(1), \cdots, \mathbf{n}_d(L)]^T$, in which $\mathbf{x}_t(l)$, $\mathbf{y}(l)$ and $\mathbf{n}_d(l)$ denote the transmitted signal, received signal and noise in the $l$th slot, respectively, and

$$\mathbf{\Omega} \triangleq \begin{bmatrix} \mathbf{A}_1 & \mathbf{0} & \cdots & \mathbf{0} \\ \mathbf{0} & \mathbf{A}_2 & \cdots & \mathbf{0} \\ \vdots & \vdots & \ddots & \vdots \\ \mathbf{0} & \mathbf{0} & \cdots & \mathbf{A}_L \end{bmatrix}, \tag{6}$$

in which $\mathbf{A}_l \triangleq \mathbf{G} \mathbf{\Theta}_l \mathbf{H}$. Hence, the matrix of the end-to-end channel is a block diagonal matrix consisting of $\mathbf{A}_l$'s. Assuming that the transmitted signals in each slot are independent of each other, the mutual information between the input and the output of (5) can be written as

$$\mathcal{I}(\mathbf{X}_t; \mathbf{Y}) = \sum_{l=1}^{L} \log \left| \mathbf{I} + \alpha^2 \frac{P}{m} \mathbf{A}_l \mathbf{A}_l^H \right|, \tag{7}$$

in which it is assumed that $\mathbf{x}_t(l) \sim \mathcal{CN}(\mathbf{0}, \frac{P}{m} \mathbf{I}_m)$, $\forall l = 1, \cdots, k$. Using the above equation, the probability of outage can be written as

$$\mathbb{P}\{\mathcal{O}\} = \mathbb{P} \left\{ \sum_{l=1}^{L} \log \left| \mathbf{I} + \alpha^2 \frac{P}{m} \mathbf{A}_l \mathbf{A}_l^H \right| < Lr \log(P) \right\}, \tag{8}$$

or equivalently,

$$\mathbb{P}\{\mathcal{O}\} = \mathbb{P} \left\{ \sum_{l=1}^{L} \sum_{j=1}^{\min(p,q)} \log \left( 1 + \alpha^2 \frac{P}{m} \lambda_j(\mathbf{A}_l) \right) < Lr \log(P) \right\}, \tag{9}$$

---

[5]Note that, as can be observed from the proof of Lemma 1 in Appendix I, the constraint $\alpha \leq 1$ does not affect the DMT of the system.

where $\lambda_i(\mathbf{A})$ denotes the $i$th ordered eigenvalue of $\mathbf{A}^H \mathbf{A}$ ($\lambda_1 > \lambda_2 > \cdots > \lambda_{\min}$). Defining $\gamma_j(\mathbf{B}) \triangleq -\frac{\log(\lambda_j(\mathbf{B}))}{\log(P)}$ and $\delta \triangleq -\frac{\log(\alpha^2)}{\log(P)}$, we have

$$
\begin{aligned}
\mathbb{P}\{\mathcal{O}\} &= \mathbb{P}\left\{\sum_{l=1}^{L}\sum_{j=1}^{\min(p,q)} \log\left(1 + \frac{1}{m} P^{1-\delta-\gamma_j(\mathbf{A}_l)}\right) < Lr\log(P)\right\} \\
&\doteq \mathbb{P}\left\{\sum_{l=1}^{L}\sum_{j=1}^{\min(p,q)} (1-\delta-\gamma_j(\mathbf{A}_l))^+ < Lr\right\}.
\end{aligned}
\tag{10}
$$

First, we show that $\alpha \doteq 1$ (or $\delta \doteq 0$), with probability one[6]. For this purpose, we write $\alpha^2$ as follows:

$$
\begin{aligned}
\alpha^2 &= \min\left(1, \frac{P}{\mathbb{E}_{\mathbf{x}_t, \mathbf{n}_r}\{\|\mathbf{H}\mathbf{x}_t + \mathbf{n}_r\|^2\}}\right) \\
&= \min\left(1, \frac{P}{\mathsf{Tr}\{\mathbf{H}\mathbf{Q}_{\mathbf{x}_t}\mathbf{H}^H + \mathbf{I}\}}\right) \\
&= \min\left(1, \frac{P}{\mathsf{Tr}\{\frac{P}{m}\mathbf{H}\mathbf{H}^H + \mathbf{I}\}}\right) \\
&= \min\left(1, \frac{P}{\frac{P}{m}\|\mathbf{H}\|^2 + m}\right).
\end{aligned}
\tag{11}
$$

From the above equation, we have

$$
\mathbb{P}\{\delta \geq \delta_0\} < \mathbb{P}\left\{\|\mathbf{H}\|^2 > P^{\delta_0-\epsilon}\right\},
\tag{12}
$$

for all $\delta_0, \epsilon > 0$. Noting that $\|\mathbf{H}\|^2$ has Chi-square distribution with $2mp$ degrees of freedom, it follows that

$$
\mathbb{P}\left\{\|\mathbf{H}\|^2 > P^{\delta_0-\epsilon}\right\} \sim \frac{P^{mp(\delta_0-\epsilon)}}{(mp)!}\exp\left\{-P^{\delta_0-\epsilon}\right\}.
\tag{13}
$$

Choosing $\epsilon = \frac{\delta_0}{2}$, the above equation implies that for $\delta_0 > 0$, the probability $\mathbb{P}\{\delta \geq \delta_0\}$ approaches to zero much faster than polynomially. More precisely, defining the event $\mathscr{F} \equiv \{\delta > 0\}$, we have $\mathbb{P}\{\mathscr{F}\} = o(P^{-c})$ for any positive constant $c$. Since $\mathbb{P}\{\mathcal{O}\} \dot{\geq} P^{-(q-r)(p-r)}$ (lower-bound on the outage probability corresponding to the DMT upper-bound ), we can write

$$
\begin{aligned}
\mathbb{P}\{\mathcal{O}\} &= \mathbb{P}\{\mathcal{O}|\mathscr{F}\}\mathbb{P}\{\mathscr{F}\} + \mathbb{P}\{\mathcal{O}|\mathscr{F}^c\}\mathbb{P}\{\mathscr{F}^c\} \\
&\stackrel{(a)}{\sim} \mathbb{P}\{\mathcal{O}|\mathscr{F}^c\},
\end{aligned}
\tag{14}
$$

where $(a)$ follows from the fact that $\mathbb{P}\{\mathscr{F}\} = o\left(\mathbb{P}\{\mathcal{O}\}\right)$. In other words, one can replace $\delta$ with zero in (10), which results in

$$
\mathbb{P}\{\mathcal{O}\} \doteq \mathbb{P}\left\{\sum_{l=1}^{L}\sum_{j=1}^{\min(p,q)} (1-\gamma_j(\mathbf{A}_l))^+ < Lr\right\}.
\tag{15}
$$

Moreover, we have

$$
\begin{aligned}
\lambda_i(\mathbf{A}_l) &\leq \|\mathbf{A}_l\|^2 \\
&\stackrel{(a)}{\leq} \|\mathbf{G}\|^2\|\mathbf{H}\|^2,
\end{aligned}
\tag{16}
$$

---

[6]Note that due to the definition of $\alpha$, we always have $\delta \geq 0$.

where $(a)$ results from the fact that $\|\mathbf{AB}\|^2 \leq \|\mathbf{A}\|^2 \|\mathbf{B}\|^2$, for any two matrices $\mathbf{A}$ and $\mathbf{B}$. Consider a negative number $\varepsilon$. From the above equation, it follows that

$$
\begin{aligned}
\mathbb{P}\{\gamma_i(\mathbf{A}_l) \leq \varepsilon\} &\leq \mathbb{P}\left\{\|\mathbf{G}\|^2 \|\mathbf{H}\|^2 \geq P^{-\varepsilon}\right\} \\
&\leq \mathbb{P}\left\{\|\mathbf{G}\|^2 \geq P^{-\frac{\varepsilon}{2}}\right\} + \mathbb{P}\left\{\|\mathbf{H}\|^2 \geq P^{-\frac{\varepsilon}{2}}\right\} \\
&\overset{(13)}{\sim} \frac{P^{-\frac{npe}{2}}}{(np)!}\exp\left\{-P^{-\frac{\varepsilon}{2}}\right\} + \frac{P^{-\frac{mpe}{2}}}{(mp)!}\exp\left\{-P^{-\frac{\varepsilon}{2}}\right\} \\
&= o\left(P^{-(q-r)(p-r)}\right).
\end{aligned}
\tag{17}
$$

As a result, following (14), we can assume that $\gamma_j(\mathbf{A}_l) \geq 0$, $\forall j = 1, \cdots, \min(p, q)$, in (15).

In order to compute the outage probability in (15), we need to find the statistical behavior of $\gamma_j(\mathbf{A}_l)$. Since we are interested in upper-bounding the outage probability of the RS scheme, finding an upper-bound for $\gamma_j(\mathbf{A}_l)$, or equivalently, a lower-bound for $\lambda_j(\mathbf{A}_l)$ would be sufficient. This is performed in the following lemma.

**Lemma 3** *Consider matrices $\mathbf{G}$ and $\mathbf{H}$ with the size of $m \times p$ and $p \times n$, respectively, and a $p \times p$ matrix $\mathbf{\Theta}$. Assume $\mathbf{G}$ and $\mathbf{H}$ are singular value decomposed as $\mathbf{G} = \mathbf{U}(\mathbf{G})\mathbf{\Lambda}^{\frac{1}{2}}(\mathbf{G})\mathbf{V}^H(\mathbf{G})$ and $\mathbf{H} = \mathbf{U}(\mathbf{H})\mathbf{\Lambda}^{\frac{1}{2}}(\mathbf{H})\mathbf{V}^H(\mathbf{H})$, respectively. We have*

$$
\lambda_i(\mathbf{G}\mathbf{\Theta}\mathbf{H}) \geq \lambda_i(\mathbf{G})\lambda_i(\mathbf{H})\lambda_{\min}\left(\mathbf{V}_{(1,i)}^H(\mathbf{G})\mathbf{\Theta}\mathbf{U}_{(1,i)}(\mathbf{H})\right),
\tag{18}
$$

*where $\lambda_i(\mathbf{A})$ and $\lambda_{\min}(\mathbf{A})$ denote the $i$'th largest eigenvalue and the minimum eigenvalue of $\mathbf{A}^H\mathbf{A}$, respectively, and $\mathbf{A}_{(a,b)}$ denotes the submatrix of $\mathbf{A}$ consisting of the $a, a+1, \ldots, b$'th columns of $\mathbf{A}$.*

*Proof:* See Appendix III. ∎

The above lemma relates $\lambda_i(\mathbf{A}_l)$ to $\lambda_i(\mathbf{G})$ and $\lambda_i(\mathbf{H})$, which facilitates the subsequent derivations. A direct consequence of the above lemmas is that

$$
\gamma_i(\mathbf{A}_l) \leq \gamma_i(\mathbf{G}) + \gamma_i(\mathbf{H}) + \gamma_{\min}(\mathbf{\Psi}_{i,l}),
\tag{19}
$$

where $\mathbf{\Psi}_{i,l} \triangleq \mathbf{V}_{(1,i)}^H(\mathbf{G})\mathbf{\Theta}_l\mathbf{U}_{(1,i)}(\mathbf{H})$. As the statistical behaviors of $\gamma_i(\mathbf{G})$ and $\gamma_i(\mathbf{H})$ are known from [16], it is sufficient to derive the asymptotic behavior of $\gamma_{\min}(\mathbf{\Psi}_{i,l})$, or equivalently, $\lambda_{\min}(\mathbf{\Psi}_{i,l})$, which is performed in the following lemma:

**Lemma 4** *Assuming small enough $\varepsilon$, we have*

$$
\mathbb{P}\{\lambda_{\min}(\mathbf{\Psi}_{i,l}) \leq \varepsilon\} \leq \eta\sqrt{\varepsilon},
\tag{20}
$$

*for some constant $\eta$.*

*Proof:* See Appendix IV. ∎

A direct consequence of the above lemma is that

$$
\mathbb{P}\{\gamma_{\min}(\mathbf{\Psi}_{i,l}) > \theta\} \overset{.}{\leq} P^{-\frac{\theta}{4}}.
\tag{21}
$$

Defining the $L \times 1$ vector $\boldsymbol{\psi} \triangleq [\psi(1), \cdots, \psi(L)]^T$ as $\psi(l) \triangleq \max_i \gamma_{\min}(\boldsymbol{\Psi}_{i,l})$, we have

$$
\begin{aligned}
\mathbb{P}\{\boldsymbol{\psi} \geq \boldsymbol{\psi}_0\} &\overset{(a)}{=} \prod_{l=1}^{L} \mathbb{P}\{\psi(l) \geq \psi_0(l)\} \\
&= \prod_{l=1}^{L} \mathbb{P}\left\{ \bigcup_{i=1}^{\min(p,q)} (\gamma_{\min}(\boldsymbol{\Psi}_{i,l}) \geq \psi_0(l)) \right\} \\
&\overset{(b)}{\leq} P^{-\frac{\mathbf{1}\cdot\boldsymbol{\psi}_0}{\min(p,q)}}
\end{aligned}
\tag{22}
$$

As $\boldsymbol{\Theta}_l$'s are independent isotropic unitary matrices, their products with any possibly correlated set of unitary matrices constructs a set of independent isotropic unitary matrices [20]. Accordingly, $\boldsymbol{\Psi}_{i,l}$'s are independent for different values of $l$, which results in $(a)$. Also, $(b)$ follows from Lemma 4 and the union bound inequality.

Let us define the $1 \times \min(p,q)$ vectors $\boldsymbol{\chi}(\mathbf{H}) \triangleq \left[\gamma_{\min(p,m)}(\mathbf{H}), \gamma_{\min(p,m)-1}(\mathbf{H}), \ldots, \gamma_{1+\min(p,m)-\min(p,q)}(\mathbf{H})\right]$ and $\boldsymbol{\chi}(\mathbf{G}) \triangleq \left[\gamma_{\min(p,n)}(\mathbf{G}), \gamma_{\min(p,n)-1}(\mathbf{G}), \ldots, \gamma_{1+\min(p,n)-\min(p,q)}(\mathbf{G})\right]$, respectively. Notice that these vectors include the log-values of the corresponding $\min(p,q)$ smallest eigenvalues of $\mathbf{H}\mathbf{H}^H$ and $\mathbf{G}\mathbf{G}^H$, respectively. Now, applying the result of Lemma 3 to (15), we can upper-bound the outage probability of the end-to-end channel as

$$
\begin{aligned}
\mathbb{P}\{\mathcal{O}\} &\overset{\cdot}{\leq} \mathbb{P}\left\{ \sum_{l=1}^{L} \sum_{i=1}^{\min(p,q)} \left(1 - \gamma_i(\mathbf{G}) - \gamma_i(\mathbf{H}) - \gamma_{\min}(\boldsymbol{\Psi}_{i,l})\right)^+ < Lr \right\} \\
&\overset{\cdot}{\leq} \mathbb{P}\left\{ \sum_{l=1}^{L} \sum_{i=1}^{\min(p,q)} \left(1 - \gamma_i(\mathbf{G}) - \gamma_i(\mathbf{H}) - \psi(l)\right)^+ < Lr \right\} \\
&\overset{(a)}{\underset{\cdot}{\leq}} \mathbb{P}\left\{ \sum_{l=1}^{L} \sum_{i=1}^{\min(p,q)} \left(1 - \gamma_{1+\min(p,n)-i}(\mathbf{G}) - \gamma_{1+\min(p,m)-i}(\mathbf{H}) - \psi(l)\right)^+ < Lr \right\} \\
&= \mathbb{P}\left\{ \sum_{l=1}^{L} \sum_{i=1}^{\min(p,q)} \left(1 - \chi_i(\mathbf{G}) - \chi_i(\mathbf{H}) - \psi(l)\right)^+ < Lr \right\},
\end{aligned}
\tag{23}
$$

where $(a)$ follows from the fact that the log-values ($\gamma_i$'s) corresponding to the smallest eigenvalues of $\mathbf{H}\mathbf{H}^H$ and $\mathbf{G}\mathbf{G}^H$ are greater than the log-values corresponding to the largest eigenvalues of these matrices. According to (22), to upper-bound the outage probability, it is sufficient to upper-bound the probability of the region of $(\boldsymbol{\psi}, \boldsymbol{\chi}(\mathbf{H}), \boldsymbol{\chi}(\mathbf{G}))$ that satisfies (23). The following lemma gives a general formula for computing such an upper-bound:

**Lemma 5** *Consider a fixed region $\mathcal{R} \subseteq [0,\infty)^n$. Assume that a uniformly continuous[7] non-negative function $f(\mathbf{x})$ ($f(\mathbf{x}) \geq 0$) is defined over $[0,\infty)^n$ such that for all $\mathbf{x} \in [0,\infty)^n$ we have $\mathbb{P}\{\mathbf{y} \geq \mathbf{x}\} \overset{\cdot}{\leq} P^{-f(\mathbf{x})}$. Then, we have*

$$
\mathbb{P}\{\mathbf{x} \in \mathcal{R}\} \overset{\cdot}{\leq} P^{-\inf_{\mathbf{x} \in \mathcal{R}} f(\mathbf{x})}.
\tag{24}
$$

*Proof:* See Appendix V. ∎

---

[7] A *uniformly continuous* function $f : \mathcal{M} \to \mathcal{N}$ where $\mathcal{M} \subseteq \mathbb{R}^m, \mathcal{N} \subseteq \mathbb{R}^n$ is a function that has the following property: for every $\epsilon$, there exists a constant $g(\epsilon) > 0$ such that for all $\mathbf{x}, \mathbf{y} \in \mathcal{M}, \|\mathbf{x} - \mathbf{y}\| \leq g(\epsilon)$, we have $\|f(\mathbf{x}) - f(\mathbf{y})\| \leq \epsilon$.

According to the upper-bound in (22) and the distribution of $\chi(\mathbf{G}), \chi(\mathbf{H})$ derived in [16], we have

$$
\begin{aligned}
\mathbb{P}\left\{\psi \geq \hat{\psi}, \chi(\mathbf{G}) \geq \chi', \chi(\mathbf{H}) \geq \chi''\right\} \quad &\dot{\leq} \quad P^{-\frac{1}{\min(p,q)}\sum_{l=1}^{L}\hat{\psi}(l)-\sum_{i=1}^{\min(p,q)}\mathsf{a}_i\chi_i''+\mathsf{b}_i\chi_i'} \\
&\overset{(a)}{\dot{\leq}} \quad P^{-\frac{1}{\min(p,q)}\sum_{l=1}^{L}\hat{\psi}(l)-\sum_{i=1}^{\min(p,q)}(2i-1+|p-q|)\left(\chi_i'+\chi_i''\right)},
\end{aligned} \tag{25}
$$

where $\mathsf{a}_i \triangleq 2[i+\min(p,m)-\min(p,q)]-1+|p-m|$ and $\mathsf{b}_i \triangleq 2[i+\min(p,n)-\min(p,q)]-1+|p-n|$ and $(a)$ follows from the fact that

$$
\begin{aligned}
\mathsf{a}_i \quad &= \quad 2[i+\min(p,m)-\min(p,q)]-1+|p-m| \\
&= \quad 2i-1+m+p-2\min(p,q) \\
&\geq \quad 2i-1+q+p-2\min(p,q) \\
&= \quad 2i-1+|p-q|,
\end{aligned} \tag{26}
$$

and similarly, $\mathsf{b}_i \geq 2i-1+|p-q|$. Now, we can apply the result of Lemma 5 to the region defined in (23) and the upper-bound derived in (25). Accordingly, we have

$$
\mathbb{P}\left\{\mathcal{O}\right\} \dot{\leq} P^{-\min_{(\chi(\mathbf{G}),\chi(\mathbf{H}),\psi)\in\mathcal{R}}\frac{1}{\min(p,q)}\sum_{l=1}^{L}\psi(l)+\sum_{i=1}^{\min(p,q)}(2i-1+|p-q|)(\chi_i(\mathbf{G})+\chi_i(\mathbf{H}))}, \tag{27}
$$

where the region $\mathcal{R}$ is defined as

$$
\begin{aligned}
\mathcal{R} \quad &\triangleq \quad \Bigg\{ (\chi(\mathbf{G}),\chi(\mathbf{H}),\psi) \Bigg| \psi \geq \mathbf{0}, \chi_1(\mathbf{G}) \geq \cdots \geq \chi_{\min(p,q)}(\mathbf{G}) \geq 0, \chi_1(\mathbf{H}) \geq \cdots \geq \chi_{\min(p,q)}(\mathbf{H}) \geq 0 \\
&, \sum_{l=1}^{L}\sum_{i=1}^{\min(p,q)}(1-\chi_i(\mathbf{G})-\chi_i(\mathbf{H})-\psi(l))^+ \leq Lr \Bigg\}.
\end{aligned} \tag{28}
$$

Let us assume $L \geq \min(p,q)\left(\sum_{i=1}^{\min(p,q)}2i-1+|p-q|\right) = \min^2(p,q)\max(p,q)$. We define $\min(p,q) \times 1$ vector $\boldsymbol{\varphi} \triangleq [\varphi_1, \cdots, \varphi_{\min(p,q)}]^T$ as $\varphi_i \triangleq \chi_i(\mathbf{G})+\chi_i(\mathbf{H})+\frac{1}{L}\sum_{l=1}^{L}\psi(l)$. For each $(\chi(\mathbf{G}),\chi(\mathbf{H}),\psi) \in \mathcal{R}$, we have

$$
\begin{aligned}
Lr \quad &\geq \quad \sum_{l=1}^{L}\sum_{i=1}^{\min(p,q)}(1-\chi_i(\mathbf{G})-\chi_i(\mathbf{H})-\psi(l))^+ \\
&= \quad \sum_{i=1}^{\min(p,q)}\sum_{l=1}^{L}\max\left\{0, 1-\chi_i(\mathbf{G})-\chi_i(\mathbf{H})-\psi(l)\right\} \\
&\geq \quad \sum_{i=1}^{\min(p,q)}\max\left\{0, \sum_{l=1}^{L}1-\chi_i(\mathbf{G})-\chi_i(\mathbf{H})-\psi(l)\right\} \\
&= \quad L\sum_{i=1}^{\min(p,q)}(1-\varphi_i)^+.
\end{aligned} \tag{29}
$$

On the other hand, according to (27) we have

$$
\begin{aligned}
\mathbb{P}\{\mathcal{O}\} \quad &\dot{\leq} \quad P^{-\min_{(\chi(\mathbf{G}),\chi(\mathbf{H}),\psi)\in\mathcal{R}}\frac{1}{\min(p,q)}\sum_{l=1}^{L}\psi(l)+\sum_{i=1}^{\min(p,q)}(2i-1+|p-q|)(\chi_i(\mathbf{G})+\chi_i(\mathbf{H}))} \\
&\dot{\leq} \quad P^{-\min_{(\chi(\mathbf{G}),\chi(\mathbf{H}),\psi)\in\mathcal{R}}\sum_{i=1}^{\min(p,q)}(2i-1+|p-q|)\varphi_i} \\
&\overset{(29)}{\dot{\leq}} \quad P^{-\min_{\boldsymbol{\varphi}\in\hat{\mathcal{R}}}\sum_{i=1}^{\min(p,q)}(2i-1+|p-q|)\varphi_i},
\end{aligned} \tag{30}
$$

where $\hat{\mathcal{R}}$ is defined as

$$\hat{\mathcal{R}} \triangleq \left\{ \boldsymbol{\varphi} \,\middle|\, \varphi_1 \geq \cdots \geq \varphi_{\min(p,q)} \geq 0, \quad \sum_{i=1}^{\min(p,q)} (1 - \varphi_i)^+ \leq r \right\}, \tag{31}$$

noting that according to the definition of $\boldsymbol{\varphi}$ we can easily conclude that $\varphi_1 \geq \cdots \geq \varphi_{\min(p,q)} \geq 0$. According to [16], (30) defines the probability of outage from the rate $r \log(P)$ in an equivalent $p \times q$ point-to-point multi-antenna Rayleigh fading channel. Hence, we have

$$d_{RS}(r) \geq d_{p \times q}(r). \tag{32}$$

On the other hand, due to the cut-set bound Theorem [21] we know that the DMT of the system is upper-bounded by the minimum of the DMT of the equivalent point-to-point $p \times m$ and $n \times p$ multi-antenna channels. Hence,

$$d_{RS}(r) \leq d_{opt}(r) = d_{p \times q}(r). \tag{33}$$

Comparing (32) and (33) completes the proof. ∎

The statement of Theorem 1 can be generalized to multi-hop networks. However, in a general multi-hop network, the RS scheme does not necessarily achieve the optimum DMT for any number of antennas at the network nodes. The following theorem gives a sufficient condition for the RS scheme to achieve the optimal DMT in a multi-hop network:

**Theorem 2** *Consider a multi-antenna multi-hop network consisting of a single source and destination and full-duplex relays, with exactly one relay in each hop. Assume that each relay is connected to the relays in the previous and next hop. Moreover, assume that for a fixed $1 \leq m \leq h$, we have $\max(N_m, N_{m-1}) \leq \min(N_0, N_1, \ldots, N_{m-2}, N_{m+1}, \ldots, N_h)$ where $h$ denotes the number of hops and $N_i$ denotes the number of antennas at the relay in the $i$'th hop. ($N_0$ and $N_h$ denote the number of antennas at the source and destination, respectively). Providing $L$ is large enough such that $L \geq \min^2(N_m, N_{m-1}) \max(N_m, N_{m-1})$, the RS scheme achieves the optimum DMT, which is the piecewise-linear function connecting the points $(k, (N_m - k)(N_{m-1} - k)), k = 0, 1, \ldots, \min(N_m, N_{m-1})$.*

*Proof:* Using the same argument as in Theorem 1, we can show that the probability of outage from the rate $r \log(P)$ is equal to

$$\mathbb{P}\{\mathcal{O}\} \doteq \mathbb{P}\left\{ \sum_{l=1}^{L} \sum_{j=1}^{N_{\min}} (1 - \gamma_j(\mathbf{A}_l))^+ < Lr \right\}, \tag{34}$$

where $N_{\min} \triangleq \min\{N_m, N_{m-1}\}$, $A_l \triangleq \mathbf{G}_h \boldsymbol{\Theta}_{l,h-1} \mathbf{G}_{h-1} \cdots \boldsymbol{\Theta}_{l,1} \mathbf{G}_1$, and $\mathbf{G}_i$ denotes the channel matrix between the nodes of the $i$'th hop and $i-1$'th hop. On the other hand, applying the argument of Lemma 3, we have

$$\gamma_i(\mathbf{A}_l) = \gamma_i(\mathbf{G}_h) + \gamma_{\min}(\boldsymbol{\Psi}_{i,l,h-1}) + \gamma_i(\mathbf{G}_{h-1}) + \cdots + \gamma_{\min}(\boldsymbol{\Psi}_{i,l,1}) + \gamma_i(\mathbf{G}_1), \tag{35}$$

where $\boldsymbol{\Psi}_{i,l,j} \triangleq \mathbf{V}_{(1,i)}^H (\mathbf{G}_{j+1}) \boldsymbol{\Theta}_{l,j} \mathbf{U}_{(1,i)} (\mathbf{G}_j \boldsymbol{\Theta}_{l,j-1} \mathbf{G}_{j-1} \cdots \boldsymbol{\Theta}_{l,1} \mathbf{G}_1)$. Moreover, we can easily check that the statement of Lemma 4 is yet valid. Hence, similar to the proof of Theorem 1, we define the $L \times 1$ vector $\boldsymbol{\psi} \triangleq [\psi(1), \cdots, \psi(L)]^T$ as

$\psi(l) \triangleq \max_{i,j} \gamma_{\min}(\mathbf{\Psi}_{i,l,j})$. We have

$$
\begin{aligned}
\mathbb{P}\left\{\psi \geq \psi_0\right\} &\overset{(a)}{=} \prod_{l=1}^{L} \mathbb{P}\left\{\psi(l) \geq \psi_0(l)\right\} \\
&= \prod_{l=1}^{L} \mathbb{P}\left\{\bigcup_{i=1}^{N_{\min}} \bigcup_{j=1}^{h} \left(\gamma_{\min}(\mathbf{\Psi}_{i,l,j}) \geq \psi_0(l)\right)\right\} \\
&\overset{(b)}{\leq} P^{-\frac{\mathbf{1} \cdot \psi_0}{N_{\min}}}.
\end{aligned}
\tag{36}
$$

As $\mathbf{\Theta}_{l,j}$'s are independent isotropic unitary matrices, their products with any possibly correlated set of unitary matrices constructs a set of independent isotropic unitary matrices [20]. Accordingly, $\mathbf{\Psi}_{i,l,j}$'s are independent for different values of $l,j$, which results in $(a)$. Also, $(b)$ follows from Lemma 4 and the union bound inequality.

Accordingly, applying (35) to (34) and rewriting inequality series of (23), we can upper-bound the outage probability as

$$
\mathbb{P}\{\mathcal{O}\} \dot{\leq} \mathbb{P}\left\{\sum_{l=1}^{L} \sum_{i=1}^{N_{\min}} \left(1 - \sum_{j=1}^{h} \chi_i(\mathbf{G}_j) - \psi(l)\right)^+ < Lr\right\},
\tag{37}
$$

where $\chi_i(\mathbf{G}_j) \triangleq \gamma_{N_{\min}+1-i}(\mathbf{G}_j)$, i.e., the reverse ordering of $\gamma_i(\mathbf{G}_j)$'s. Let us define the $N_{\min} \times 1$ vectors $\boldsymbol{\chi}(\mathbf{G}_j)$'s as $\boldsymbol{\chi}(\mathbf{G}_j) \triangleq [\chi_1(\mathbf{G}_j), \chi_2(\mathbf{G}_j), \dots, \chi_{N_{\min}}(\mathbf{G}_j)]^T$ containing the corresponding log-values of the $N_{\min}$ smallest eigenvalues of $\mathbf{G}_j \mathbf{G}_j^H$. Notice that $\chi_1(\mathbf{G}_j) \geq \chi_2(\mathbf{G}_j) \geq \dots \geq \chi_{N_{\min}}(\mathbf{G}_j) \geq 0$. According to the upper-bound in (36) and the statistical behavior of the eigenvalues of $\mathbf{G}_j \mathbf{G}_j^H$ derived in [16], we have

$$
\begin{aligned}
\mathbb{P}\left\{\psi \geq \hat{\psi}, \boldsymbol{\chi}(\mathbf{G}_j) \geq \hat{\boldsymbol{\chi}}(\mathbf{G}_j), j = 1, \dots, h\right\} &\dot{\leq} P^{-\frac{1}{N_{\min}} \sum_{l=1}^{L} \hat{\psi}(l) - \sum_{j=1}^{h} \sum_{i=1}^{N_{\min}} (2(i+\min(N_j, N_{j-1}) - N_{\min}) - 1 + |N_j - N_{j-1}|)\hat{\chi}_i(\mathbf{G}_j)} \\
&\overset{(a)}{\leq} P^{-\frac{1}{N_{\min}} \sum_{l=1}^{L} \hat{\psi}(l) - \sum_{i=1}^{N_{\min}} (2i - 1 + |N_m - N_{m-1}|)\left(\sum_{j=1}^{h} \hat{\chi}_i(\mathbf{G}_j)\right)}.
\end{aligned}
\tag{38}
$$

Here, $(a)$ results from the fact that $2\min(N_j, N_{j-1}) - 2N_{\min} + |N_j - N_{j-1}| = N_j + N_{j-1} - 2N_{\min} \geq N_m + N_{m-1} - 2N_{\min} = |N_m - N_{m-1}|$ which comes form the assumption of $\max(N_m, N_{m-1}) \leq \min(N_0, N_1, \dots, N_{m-2}, N_{m+1}, \dots, N_h)$. Now, we can apply the result of Lemma 5 to the region defined in (38) and the upper-bound derived in (37). Accordingly, we have

$$
\mathbb{P}\{\mathcal{O}\} \dot{\leq} P^{-\min_{(\boldsymbol{\chi}(\mathbf{G}_1), \dots, \boldsymbol{\chi}(\mathbf{G}_h), \psi) \in \mathcal{R}} \frac{1}{N_{\min}} \sum_{l=1}^{L} \psi(l) + \sum_{i=1}^{N_{\min}} (2i - 1 + |N_m - N_{m-1}|)\left(\sum_{j=1}^{h} \hat{\chi}_i(\mathbf{G}_j)\right)},
\tag{39}
$$

where the region $\mathcal{R}$ is defined as

$$
\mathcal{R} \triangleq \left\{ (\boldsymbol{\chi}(\mathbf{G}_1), \boldsymbol{\chi}(\mathbf{G}_2), \dots, \boldsymbol{\chi}(\mathbf{G}_h), \psi) \,\middle|\, \psi \geq \mathbf{0}, \chi_1(\mathbf{G}_j) \geq \chi_2(\mathbf{G}_j) \geq \dots \geq \chi_{\min(p,q)}(\mathbf{G}_j) \geq 0, j = 1, 2, \dots, h \right.
$$
$$
\left. , \sum_{l=1}^{L} \sum_{i=1}^{\min(p,q)} \left(1 - \psi(l) - \sum_{j=1}^{h} \chi_i(\mathbf{G}_j)\right)^+ \leq Lr \right\}.
\tag{40}
$$

Similar to the proof of Theorem 1, we define $N_{\min} \times 1$ vector $\boldsymbol{\varphi} \triangleq [\varphi_1, \dots, \varphi_{N_{\min}}]^T$ as $\varphi_i \triangleq \sum_{j=1}^{h} \chi_i(\mathbf{G}_j) + \frac{1}{L} \sum_{l=1}^{L} \psi(l)$. Rewriting the inequality series in (29) and (30), we can upper-bound the outage probability as

$$
\mathbb{P}\{\mathcal{O}\} \dot{\leq} P^{-\min_{\boldsymbol{\varphi} \in \hat{\mathcal{R}}} \sum_{i=1}^{N_{\min}} (2i - 1 + |N_m - N_{m-1}|)\varphi_i},
\tag{41}
$$

where $\hat{\mathcal{R}}$ is defined as

$$
\hat{\mathcal{R}} \triangleq \left\{ \boldsymbol{\varphi} \,\middle|\, \varphi_1 \geq \dots \geq \varphi_{N_{\min}} \geq 0, \sum_{i=1}^{\min(p,q)} (1 - \varphi_i)^+ \leq r \right\}.
\tag{42}
$$

According to [16], (41) and (42) define the probability of outage from the rate $r \log(P)$ in an equivalent $N_m \times N_{m-1}$ point-to-point multi-antenna Rayleigh fading channel. Hence, we have

$$d_{RS}(r) \geq d_{N_m \times N_{m-1}}(r). \tag{43}$$

On the other hand, due to the cut-set bound Theorem [21], we know that the DMT of the system is upper-bounded by the minimum of the DMT of the channels of different hops. Hence,

$$d_{RS}(r) \leq d_{opt}(r) = d_{N_m \times N_{m-1}}(r). \tag{44}$$

Comparing (43) and (44) completes the proof. ∎

**Corollary 1** *Consider a multi-antenna multi-hop network consisting of a single source, a single destination and full-duplex relays with exactly one relay in each hop and assume that all the nodes are equipped with $N$ antennas. Providing $L$ is large enough such that $L \geq N^3$, the RS scheme achieves the optimum DMT, which is the piecewise-linear function connecting the points $(k, (N-k)^2), k = 0, 1, \ldots, N$.*

### B. Parallel Relay Network

In this subsection, we consider the setup of a multi-antenna parallel relay network. In specific, we consider a two-hop network consisting of $K > 1$ half-duplex relays with the assumption that there is no direct link between the source and the destination. The source and the destination are shown by nodes $0$ and $K + 1$, respectively, while the $K$ parallel relays are denoted by the nodes $1, 2, \ldots, K$. Earlier, the optimum DMT of the single-antenna parallel relay network is derived by [1] and independently by [18]. Indeed, it is shown in [1] that the RS scheme can achieve the optimum DMT of the single-antenna multiple-access parallel relay network. However, much less is known regarding the DMT of the multi-antenna parallel relay networks.

Here, we show that the RS scheme achieves a better DMT with respect to the traditional AF relaying and also with respect to the other results reported in the literature. Moreover, we show that the RS scheme achieves the optimum DMT of the 2-relay parallel relay network.

**Theorem 3** *Consider a multi-antenna parallel relay network consisting of a source equipped with $m$ antennas, a destination equipped with $n$ antennas and $K$ half-duplex relays each equipped with $p$ antennas. Assume that there exists no direct link between the source and the destination[8]. For any fixed $B \geq \min^2(p, q) \max(p, q)$, the RS scheme[9] with $L = BK$ number of paths, $S = BK + 1$ number of slots, the path sequence*

$$Q \equiv (q_1, \ldots, q_K, q_1, \ldots, q_K, \ldots, q_1, \ldots, q_K),$$

*in which $q_k \equiv (0, k, K + 1)$, and the timing sequence $s_{i,j} = i + j - 1$, achieves the diversity gain*

$$d_{RS}(r) \geq K d_{p \times q}\left(\left(1 + \frac{1}{BK}\right)r\right), \tag{45}$$

---

[8]Note that in this theorem, the relays are not assumed to be isolated from each other, i.e., there may exist some links between the relays.

[9]The reader is encouraged to have a look at the definition of RS scheme and its parameters in [1].

*where $q \triangleq \min(m,n)$ and $d_{p \times q}(r)$ denotes the diversity gain of the point-to-point $p \times q$ multi-antenna Rayleigh fading channel corresponding to the rate $r \log(P)$. Moreover, as $B \to \infty$, the RS scheme achieves the diversity gain $K d_{p \times q}(r)$.*

*Proof:* The proof steps are the same as the ones presented for Theorem 3 of [1] and Theorem 6.7 of [18]. Let us denote the channel between the $k$'th relay and the source and the channel between the $k$'th relay and the destination by $\mathbf{H}_k$ and $\mathbf{G}_k$, respectively. Moreover, let us define $r(i) \triangleq (i-1) \mod K + 1$. Similar to the proof of Theorem 3 in [1], one can easily show that the end-to-end channel from the source to the destination can be shown by a block lower-triangular matrix. More precisely, we have

$$\mathbf{y} = \mathcal{F}\mathbf{x} + \mathcal{Q}\mathbf{n}_r + \mathbf{n}_d. \tag{46}$$

Here, $\mathbf{x}$ denotes the vector corresponding to all the paths transmitted by the source, $\mathbf{y}$ denotes the vector corresponding to all the paths received by the destination,

$$\mathcal{F} = \begin{pmatrix} \mathbf{F}_{1,1} & \mathbf{0} & \mathbf{0} & \dots \\ \mathbf{F}_{2,1} & \mathbf{F}_{2,2} & \mathbf{0} & \dots \\ \vdots & \vdots & \vdots & \ddots \\ \mathbf{F}_{L,1} & \mathbf{F}_{L,2} & \dots & \mathbf{F}_{L,L} \end{pmatrix}, \tag{47}$$

where[10] $\mathbf{F}_{i,i} = \mathbf{G}_{r(i)} \alpha_i \mathbf{\Theta}_i \mathbf{H}_{r(i)}$, and

$$\mathcal{Q} = \begin{pmatrix} \mathbf{Q}_{1,1} & \mathbf{0} & \mathbf{0} & \dots \\ \mathbf{Q}_{2,1} & \mathbf{Q}_{2,2} & \mathbf{0} & \dots \\ \vdots & \vdots & \vdots & \ddots \\ \mathbf{Q}_{L,1} & \mathbf{Q}_{L,2} & \dots & \mathbf{Q}_{L,L} \end{pmatrix}, \tag{48}$$

where $\mathbf{Q}_{i,i} = \mathbf{G}_{r(i)} \alpha_i \mathbf{\Theta}_i$. The DMT of the end-to-end channel is equal to

$$
\begin{aligned}
d_{RS}(r) &= \lim_{P \to \infty} -\frac{\log\left(\mathbb{P}\left\{I(\mathbf{x};\mathbf{y}) < (L+1)r\log(P)\right\}\right)}{\log(P)} \\
&= \lim_{P \to \infty} -\frac{\log\left(\mathbb{P}\left\{\log\left(\left|\mathbf{I}_{Ln} + P\mathcal{F}\mathcal{F}^H \left(\mathbf{I}_{Ln} + \mathcal{Q}\mathcal{Q}^H\right)^{-1}\right|\right) < (L+1)r\log(P)\right\}\right)}{\log(P)}.
\end{aligned}
\tag{49}
$$

Noting $\mathcal{F}$ is block lower-diagonal and applying Theorem 3.3 in [18], we have

$$\left|\mathbf{I}_{Ln} + P\mathcal{F}\mathcal{F}^H \left(\mathbf{I}_{Ln} + \mathcal{Q}\mathcal{Q}^H\right)^{-1}\right| \geq \prod_{l=1}^{L} \left|\mathbf{I}_n + P\mathbf{F}_{l,l}\mathbf{F}_{l,l}^H \left(\mathbf{I}_n + \sum_{i=1}^{l} \mathbf{Q}_{l,i}\mathbf{Q}_{l,i}^H\right)^{-1}\right|. \tag{50}$$

Note that according to the constraint in the RS scheme, we have $\alpha_l \leq 1$. Hence, one can apply the same argument as in Lemma 2 and show that

$$\mathbb{P}\left\{\prod_{l=1}^{L} \left|\mathbf{I}_n + P\mathbf{F}_{l,l}\mathbf{F}_{l,l}^H \left(\mathbf{I}_n + \sum_{i=1}^{l} \mathbf{Q}_{l,i}\mathbf{Q}_{l,i}^H\right)^{-1}\right| < P^{(L+1)r}\right\} \doteq \mathbb{P}\left\{\prod_{l=1}^{L} \left|\mathbf{I}_n + P\mathbf{F}_{l,l}\mathbf{F}_{l,l}^H\right| < P^{(L+1)r}\right\}. \tag{51}$$

---

[10] As only the value of $\mathbf{F}_{i,i}$ is needed in the proof of Theorem 3, we just give the value of $\mathbf{F}_{i,i}$ here. The formula for $\mathbf{F}_{i,j}$, $i < j$, is much more complicated and hence, we decide not to bring it. The same thing for $\mathbf{Q}_{i,j}$ defined right after this.

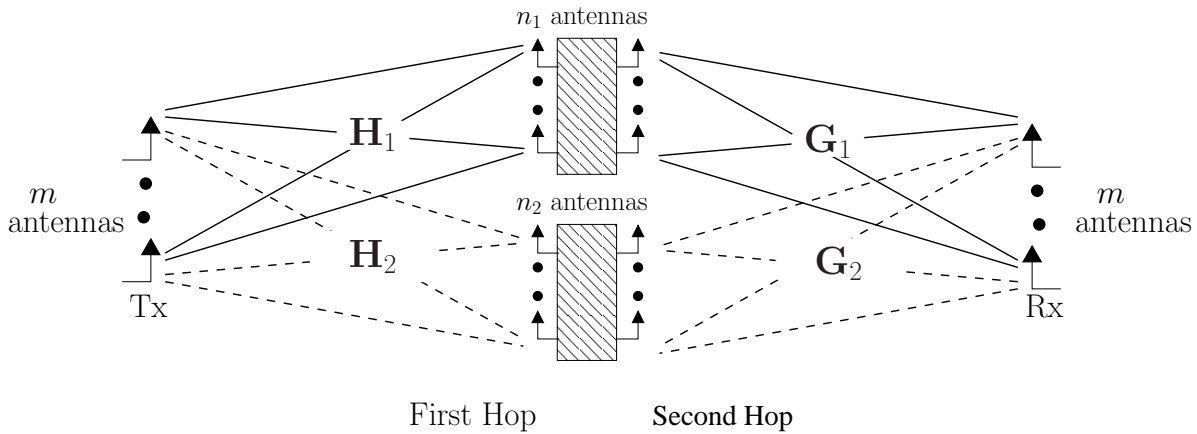

Fig. 2. A schematic of the MIMO parallel 2 relays network with no direct link between the source and the destination and also between the relays

Moreover, using the argument in the proof of Theorem 1, one can show that with probability one, we have $\alpha_l \doteq 1$. Hence, defining $\mathbf{A}_{k,b} \triangleq \mathbf{G}_k \mathbf{\Theta}_{(b-1)K+k} \mathbf{H}_k$, we have

$$\mathbb{P}\left\{\prod_{l=1}^{L}\left|\mathbf{I}_n + P\mathbf{F}_{l,l}\mathbf{F}_{l,l}^H\right| < P^{(L+1)r}\right\} \doteq \mathbb{P}\left\{\prod_{k=1}^{K}\prod_{b=1}^{B}\left|\mathbf{I}_n + P\mathbf{A}_{k,b}\mathbf{A}_{k,b}^H\right| < P^{(L+1)r}\right\}. \tag{52}$$

Let us define the $1 \times K$ random vector $\boldsymbol{\sigma} = [\sigma_1, \sigma_2, \ldots, \sigma_K]$ where $\sigma_k \triangleq \frac{\sum_{b=1}^{B} \log\left|\mathbf{I}_n + P\mathbf{A}_{k,b}\mathbf{A}_{k,b}^H\right|}{B \log(P)}$. Notice that $\sigma_k$'s are independent of each other. As $B \geq \min^2(p,q) \max(p,q)$, we can apply Theorem 1 to $\sigma_k$'s. Hence, for any fixed $1 \times K$ vector $\hat{\boldsymbol{\sigma}} \geq \mathbf{0}$, we have

$$\mathbb{P}\left\{\boldsymbol{\sigma} \geq \hat{\boldsymbol{\sigma}}\right\} \overset{(a)}{\doteq} P^{-\sum_{k=1}^{K} d_{p\times q}(\hat{\sigma}_k)}. \tag{53}$$

Here, $(a)$ results from Theorem 1 and the fact that $\sigma_k$'s are independent of each other. Denoting the outage event as $\mathcal{O}$, according to (49), (50), (51), and (52), we have

$$\mathbb{P}\left\{\mathcal{O}\right\} \dot{\leq} \mathbb{P}\left\{\sum_{k=1}^{K} \sigma_k \leq \left(K + \frac{1}{B}\right) r\right\}. \tag{54}$$

Let us define the region $\mathcal{R} \triangleq \left\{\boldsymbol{\sigma} \geq \mathbf{0} \,\middle|\, \sum_{k=1}^{K} \sigma_k \leq \left(K + \frac{1}{B}\right) r\right\}$. We have

$$\mathbb{P}\left\{\mathcal{O}\right\} \dot{\leq} \mathbb{P}\left\{\mathcal{R}\right\} \overset{(a)}{\leq} P^{-\min_{\boldsymbol{\sigma} \in \mathcal{R}} \sum_{k=1}^{K} d_{p\times q}(\sigma_k)} \overset{(b)}{=} P^{-K d_{p\times q}\left(\left(1 + \frac{1}{BK}\right)r\right)}. \tag{55}$$

Here, $(a)$ results from Lemma 5. $(b)$ results from the fact that $d_{p\times q}(r)$ is a convex decreasing function and, as a result, we have $\frac{1}{K}\sum_{k=1}^{K} d_{p\times q}(\sigma_k) \geq d_{p\times q}(\frac{1}{K}\sum_{k=1}^{K} \sigma_k) \geq d_{p\times q}\left(\left(1 + \frac{1}{BK}\right)r\right)$. (55) completes the proof of the Theorem. ∎

In the following Theorem, we show that the RS scheme achieves the optimum DMT for the two-relays half-duplex parallel relay network in which $m = n$, but the two parallel relays can have a different number of antennas but there is no direct link between them. Figure 2 shows the schematic of such a network.

**Theorem 4** *Consider a multi-antenna parallel relay network consisting of a source and a destination each equipped with $m$ antennas, and $K = 2$ half-duplex relays equipped with $n_k, k = 1, 2$ antennas. Assume that there exists no direct link between the source and the destination and also between the two relays. Consider the RS scheme with $L = BK$, $S = BK + 1$, and the path and timing sequences defined in Theorem 3. As $B \to \infty$, the RS scheme achieves the optimum DMT of the network.*

*Proof:* First, notice that according to the argument of Theorem 3, as $B \to \infty$, the RS scheme achieves the DMT $d_{RS,\infty}(r) \triangleq \min_{0 \leq \nu \leq 2r} d_{m \times n_1}(\nu) + d_{m \times n_2}(2r - \nu)$. Now, to prove the Theorem, we just have to show that $d_{RS,\infty}(r)$ is indeed an upper-bound for the optimum DMT. According to the cut-set Theorem [21], we have an upper-bound for the capacity of the network for each channel realization. Hence, we can apply the cut-set Theorem to find an upper-bound for the optimum DMT. In general, for any arbitrary half-duplex relay network with $K$ relays and any set $\{0\} \subseteq \mathcal{S} \subseteq \{0, 1, \ldots, K\}$, we say the network is in the *state* $\mathcal{S}$, if the network nodes in $\mathcal{S}$ are transmitting and the network nodes in $\mathcal{S}^c \triangleq \{0, 1, \ldots, K+1\} / \mathcal{S}$ are receiving. Notice that as the source is always transmitting and the destination is always receiving, we have $0 \in \mathcal{S}, K+1 \in \mathcal{S}^c$. Accordingly, we define a $1 \times 2^K$ *state vector* $\boldsymbol{\rho}$ such that for any set $\{0\} \subseteq \mathcal{S} \subseteq \{0, 1, \ldots, K\}$, $\rho_{\mathcal{S}}$ shows the portion of time that the half-duplex relay network spends in the *state* $\mathcal{S}$ ($\sum_{\{0\} \subseteq \mathcal{S} \subseteq \{0,1,\ldots,K\}} \rho_{\mathcal{S}} = 1$). As the channels are assumed to be fixed for the whole transmission period and the relay nodes and the source are assumed to have no channel state knowledge about their forward channels, we can assume that a fixed state vector $\boldsymbol{\rho}$ is associated with the strategy that achieves the optimum DMT. Denoting the outage event by $\mathcal{O}$, for any general half-duplex relay network consisting of $K$ relays, we have

$$
\begin{aligned}
\mathbb{P}\{\mathcal{O}\} &\overset{(a)}{\geq} \min_{\boldsymbol{\rho}} \mathbb{P}\left\{ \bigcup_{\{0\} \subseteq \mathcal{T} \subseteq \{0,1,\ldots,K\}} \left( \sum_{\{0\} \subseteq \mathcal{S} \subseteq \{0,1,\ldots,K\}} \rho_{\mathcal{S}} I\left( X\left(\mathcal{S} \cap \mathcal{T}\right); Y\left(\mathcal{S}^c \cap \mathcal{T}^c\right) \middle| X\left(\mathcal{S} \cap \mathcal{T}^c\right) \right) < r\log(P) \right) \right\} \\
&\overset{(b)}{=} \min_{\boldsymbol{\rho}} \max_{\{0\} \subseteq \mathcal{T} \subseteq \{0,1,\ldots,K\}} \mathbb{P}\left\{ \sum_{\{0\} \subseteq \mathcal{S} \subseteq \{0,1,\ldots,K\}} \rho_{\mathcal{S}} I\left( X\left(\mathcal{S} \cap \mathcal{T}\right); Y\left(\mathcal{S}^c \cap \mathcal{T}^c\right) \middle| X\left(\mathcal{S} \cap \mathcal{T}^c\right) \right) < r\log(P) \right\}. \quad (56)
\end{aligned}
$$

Here, $(a)$ follows from the cut-set bound Theorem [21] and $(b)$ follows from the union bound on the probability. Now, in our two-relay parallel setup, let us define two sets $\mathcal{T}_1 \triangleq \{0, 1\}$ and $\mathcal{T}_2 \triangleq \{0, 2\}$ corresponding to the two cut-sets. Moreover, let us define two events $\mathcal{O}_1$ and $\mathcal{O}_2$ as $\mathcal{O}_1 \triangleq \left\{ \left| \mathbf{I}_m + P\mathbf{G}_1 \mathbf{G}_1^H \right| \leq \hat{\nu}\log(P), \left| \mathbf{I}_{n_2} + P\mathbf{H}_2 \mathbf{H}_2^H \right| \leq (2r - \hat{\nu})\log(P) \right\}$ and $\mathcal{O}_2 \triangleq \left\{ \left| \mathbf{I}_{n_1} + P\mathbf{H}_1 \mathbf{H}_1^H \right| \leq \hat{\nu}\log(P), \left| \mathbf{I}_m + P\mathbf{G}_2 \mathbf{G}_2^H \right| \leq (2r - \hat{\nu})\log(P) \right\}$ where $\hat{\nu} \triangleq \underset{0 \leq \nu \leq 2r}{\operatorname{argmin}} \, d_{m \times n_1}(\nu) + d_{m \times n_2}(2r - \nu)$. Hence, in our setup, (56) can be simplified as

$$
\begin{aligned}
\mathbb{P}\{\mathcal{O}\} &\overset{(a)}{\geq} \min_{\boldsymbol{\rho}} \max \left( \mathbb{P}\left\{ \sum_{\{0\} \subseteq \mathcal{S} \subseteq \{0,1,2\}} \rho_{\mathcal{S}} I\left( X\left(\mathcal{S} \cap \mathcal{T}_1\right); Y\left(\mathcal{S}^c \cap \mathcal{T}_1^c\right) \middle| X\left(\mathcal{S} \cap \mathcal{T}_1^c\right) \right) \leq r\log(P) \right\}, \right. \\
&\qquad\qquad\qquad \left. \mathbb{P}\left\{ \sum_{\{0\} \subseteq \mathcal{S} \subseteq \{0,1,2\}} \rho_{\mathcal{S}} I\left( X\left(\mathcal{S} \cap \mathcal{T}_2\right); Y\left(\mathcal{S}^c \cap \mathcal{T}_2^c\right) \middle| X\left(\mathcal{S} \cap \mathcal{T}_2^c\right) \right) \leq r\log(P) \right\} \right) \\
&\geq \min_{\boldsymbol{\rho}} \max \left( \mathbb{P}\left\{ \left(\rho_{\{0,1\}} + \rho_{\{0,1,2\}}\right) \left| \mathbf{I}_m + P\mathbf{G}_1 \mathbf{G}_1^H \right| + \left(\rho_{\{0\}} + \rho_{\{0,1\}}\right) \left| \mathbf{I}_{n_2} + P\mathbf{H}_2 \mathbf{H}_2^H \right| \leq r\log(P) \right\}, \right. \\
&\qquad\qquad\qquad \left. \mathbb{P}\left\{ \left(\rho_{\{0,2\}} + \rho_{\{0,1,2\}}\right) \left| \mathbf{I}_m + P\mathbf{G}_2 \mathbf{G}_2^H \right| + \left(\rho_{\{0\}} + \rho_{\{0,2\}}\right) \left| \mathbf{I}_{n_1} + P\mathbf{H}_1 \mathbf{H}_1^H \right| \leq r\log(P) \right\} \right) \\
&\overset{(b)}{\geq} \min_{\boldsymbol{\rho}} \max \left( \mathbf{1}\left[ r - \left(\rho_{\{0,1,2\}} + \rho_{\{0,1\}}\right) \hat{\nu} - \left(\rho_{\{0,1\}} + \rho_{\{0\}}\right)(2r - \hat{\nu}) \right] \mathbb{P}\{\mathcal{O}_1\}, \right. \\
&\qquad\qquad\qquad \left. \mathbf{1}\left[ r - \left(\rho_{\{0\}} + \rho_{\{0,2\}}\right) \hat{\nu} - \left(\rho_{\{0,2\}} + \rho_{\{0,1,2\}}\right)(2r - \hat{\nu}) \right] \mathbb{P}\{\mathcal{O}_2\} \right) \\
&\overset{(c)}{=} P^{-d_{RS,\infty}(r)}, \quad (57)
\end{aligned}
$$

where $\mathbf{1}[x] = 1$ for $x \geq 0$ and is 0 otherwise. Here, $(a)$ results from taking the maximization of the right-hand side of (56) over $\mathcal{T}_1, \mathcal{T}_2$. $(b)$ results from the facts that i) conditioned on $\mathcal{O}_1$ and assuming $r \geq \left(\rho_{\{0,1,2\}} + \rho_{\{0,1\}}\right) \hat{\nu} + \left(\rho_{\{0,1\}} + \rho_{\{0\}}\right)(2r - \hat{\nu})$, we have $\left(\rho_{\{0,1\}} + \rho_{\{0,1,2\}}\right) \left| \mathbf{I}_m + P\mathbf{G}_1 \mathbf{G}_1^H \right| + \left(\rho_{\{0\}} + \rho_{\{0,1\}}\right) \left| \mathbf{I}_{n_2} + P\mathbf{H}_2 \mathbf{H}_2^H \right| \leq r\log(P)$; and ii) conditioned on $\mathcal{O}_2$ and assuming $r \geq \left(\rho_{\{0\}} + \rho_{\{0,2\}}\right) \hat{\nu} + \left(\rho_{\{0,2\}} + \rho_{\{0,1,2\}}\right)(2r - \hat{\nu})$, we conclude that $\left(\rho_{\{0,2\}} + \rho_{\{0,1,2\}}\right) \left| \mathbf{I}_m + P\mathbf{G}_2 \mathbf{G}_2^H \right| +$ $\left(\rho_{\{0\}} + \rho_{\{0,2\}}\right) \left|\mathbf{I}_{n_1} + P\mathbf{H}_1\mathbf{H}_1^H\right| \leq r \log(P)$. (c) results from i) $\mathbb{P}\left\{\mathcal{O}_1\right\} = \mathbb{P}\left\{\mathcal{O}_2\right\} \doteq P^{-d_{m \times n_1}(\hat{\nu}) - d_{m \times n_2}(2r - \hat{\nu})} = P^{-d_{RS,\infty}(r)}$ and ii) $\rho_0 + \rho_{0,1} + \rho_{0,2} + \rho_{0,1,2} = 1$ (due to the definition of $\boldsymbol{\rho}$) which results in having $r - \left(\rho_{\{0,1,2\}} + \rho_{\{0,1\}}\right) \hat{\nu} - \left(\rho_{\{0,1\}} + \rho_{\{0\}}\right) (2r - \hat{\nu}) = -\left[r - \left(\rho_{\{0\}} + \rho_{\{0,2\}}\right) \hat{\nu} - \left(\rho_{\{0,2\}} + \rho_{\{0,1,2\}}\right) (2r - \hat{\nu})\right]$ and consequently,

$$\mathbf{1}\left[r - \left(\rho_{\{0,1,2\}} + \rho_{\{0,1\}}\right) \hat{\nu} - \left(\rho_{\{0,1\}} + \rho_{\{0\}}\right) (2r - \hat{\nu})\right] + \mathbf{1}\left[r - \left(\rho_{\{0\}} + \rho_{\{0,2\}}\right) \hat{\nu} - \left(\rho_{\{0,2\}} + \rho_{\{0,1,2\}}\right) (2r - \hat{\nu})\right] = 1.$$

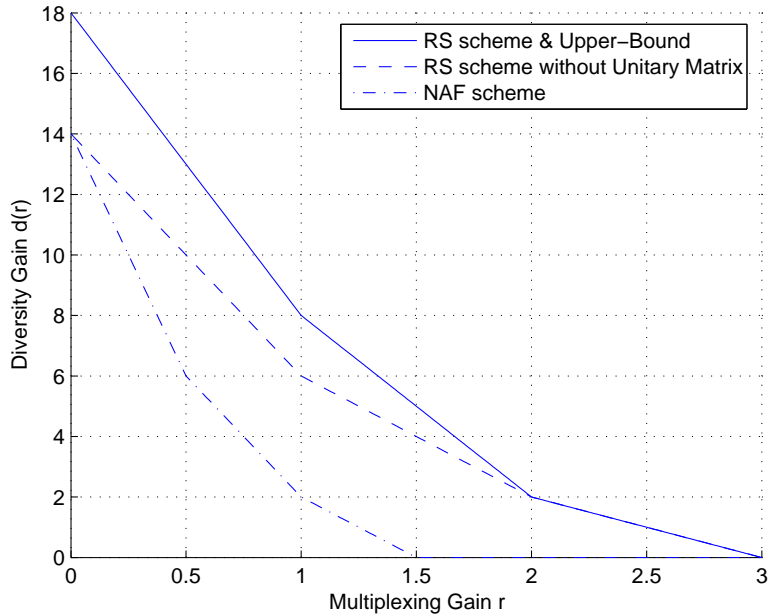

Fig. 3. Parallel relay network with $K = 2$ relays, each node with 3 antennas and no direct link between source and destination.

(57) completes the proof of the theorem. ∎

Figure 3 shows the DMT of various schemes for the parallel relay network with $K = 2$ relays and $m = n = p = 3$, i.e. 3 antennas at each node. As it is shown in Theorem 4, the RS scheme achieves the optimum DMT. However, if we do not apply random unitary matrix multiplication at the relay nodes, applying the steps in the proof of Theorem 3, one can easily show that the RS scheme achieves the DMT of $Kd_{\mathbf{GH}}(r)$, where $d_{\mathbf{GH}}(r)$ denotes the DMT of the product of the channel matrix $\mathbf{H}$ from the source to the relay and the channel matrix $\mathbf{G}$ from the relay to the destination (see (4)). Finally, applying the NAF scheme of [2], [17], one can easily show that the DMT $Kd_{\mathbf{GH}}(2r)$ is achievable.

## C. Multiple-Antenna Single Relay Channel

In this subsection, we consider the most-studied scenario in the relay network, the single relay setup, in which a direct link exists between the source and the destination. The relay is assumed to be half-duplex. There have been extensive research on this particular setup toward characterization of the DMT. The authors of [2] have shown that the NAF scheme achieves the best DMT among all possible AF relaying schemes for the Single-Input Single-Output (SISO) single half-duplex relay channel. However, here we show that using independent uniformly random unitary matrices across different time-slots improves the DMT of the NAF scheme for the multi-antenna setup. In order to exploit the potential benefit from random unitary matrix multiplication, the source transmits in $2B$ consecutive time-slots. In the odd time-slots, the relay listens to the source signal. In the even time slots, the relay multiplies the received signal from the last time-slot with a uniformly random unitary matrix and then amplifies the result with the maximum possible coefficient, which is less than or equal to 1. The destination decodes the transmitted message based on the joint decoding of the signal it receives in the $2B$ time-slots.

**Theorem 5** *Consider a single relay channel consisting of a source, a half-duplex relay, and a destination equipped with $m$, $p$, and $n$ antennas, respectively. Let us consider a modified NAF scheme that benefits from the random unitary matrix multiplication at the relay node and the joint decoding at the destination side through $2B$ time-slots. Assuming $B \geq \min^2(p, q) \max(p, q)$ where $q \triangleq \min(m, n)$, the modified NAF scheme achieves the following DMT*

$$d_{MNAF}(r) \geq d_{m \times n}(r) + d_{p \times q}(2r). \tag{58}$$

*Proof:* The proof is similar to the proof of Theorem 3. Indeed, assuming the source-destination, source-relay, and relay-destination channel matrices are denoted by $\mathbf{F}$, $\mathbf{H}$, and $\mathbf{G}$, respectively, we can show that the end-to-end channel matrix is equal to

$$\mathcal{F} = \begin{pmatrix} \mathbf{F} & \mathbf{0} & \mathbf{0} & \mathbf{0} & \dots \\ \alpha\mathbf{G}\mathbf{\Theta}_1\mathbf{H} & \mathbf{F} & \mathbf{0} & \mathbf{0} & \dots \\ \mathbf{0} & \mathbf{0} & \mathbf{F} & \mathbf{0} & \dots \\ \mathbf{0} & \mathbf{0} & \alpha\mathbf{G}\mathbf{\Theta}_2\mathbf{H} & \mathbf{F} & \dots \\ \vdots & \vdots & \vdots & \vdots & \ddots \end{pmatrix}. \tag{59}$$

Here, we observe that the lower-diagonal elements are independent of the diagonal elements. Hence, we can apply Theorem 3.3 in [18]. Accordingly, the DMT corresponding to the end-to-end system is greater than or equal to the summation of the DMT of the forward channel $\mathbf{F}$ and the DMT of a two-hop channel utilized in half of the time. In other words, $d_{MNAF}(r) \geq d_{m \times n}(r) + d_{p \times q}(2r)$. Details of the proof are similar to the proof of Theorem 3. ∎

Figure 4 compares the achievable DMT of the NAF scheme with the achievable DMT of the modified NAF scheme for the single relay channel with $m = n = p = 3$ and $m = n = p = 4$ antennas. Reference [13] has shown that the NAF protocol achieves the DMT $d_{NAF}(r) \geq d_{m \times n}(r) + d_{\mathbf{GH}}(2r)$. As we observe, the modified NAF scheme outperforms the NAF scheme for small values of $r$.

### D. General Full-Duplex Relay Networks

Here, we generalize the statement of *Remark 6* in [1] and Theorem 4.2 in [18] to the multi-antenna case. Indeed, it is shown in [1] that the RS scheme achieves a linear DMT connecting the points $(0, d_{\max})$ and $(r_{\max}, 0)$, where $d_{\max}$ denotes the maximum diversity and $r_{\max}$ denotes the maximum multiplexing-gain (which is 1) for single-antenna full-duplex relay networks whose underlying graph is directed acyclic[11]. However, in the multi-antenna setup, [1] could only show that the RS scheme achieves the maximum diversity gain. To generalize the statement, we have to review the definitions from [1].

---

[11]A directed graph is called directed acyclic if it contains no directed cycle [22]

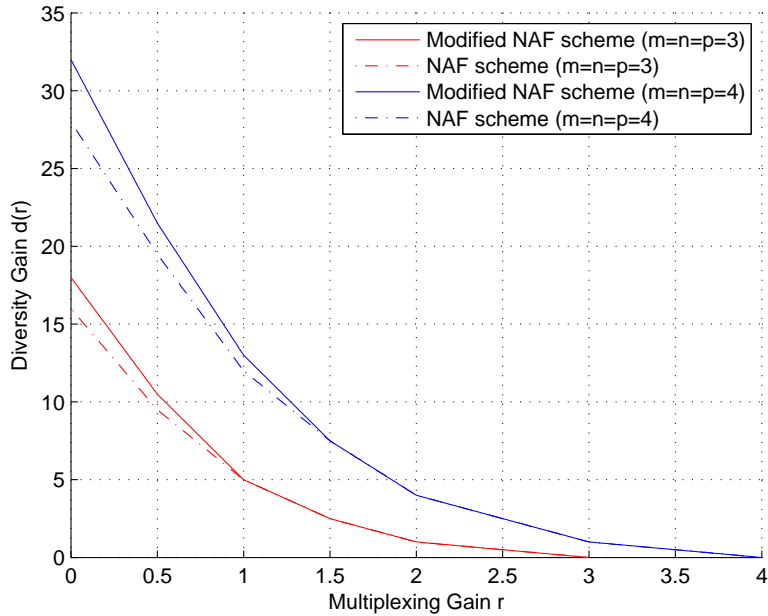

Fig. 4. DMT of the NAF scheme versus the modified NAF scheme for Multiple-antenna single relay channel with i)3 antennas ii)4 antennas at each node.

**Definition 1** *A general multi-relay network with the connectivity graph* $G = (V, E)$ *is a wireless network with the vertex set* $V = \{0, 1, \ldots, K+1\}$ *where* $0$ *and* $K+1$ *represent the source and the destination, respectively, and the other nodes represent the relays. Each pair of vertices that belong to* $E$ *are connected together through a quasi-static Rayleigh-fading channel and the pairs of vertices that do not lie in* $E$ *are disconnected from each other. The number of antennas at node* $i$ *is denoted by* $N_i$.

**Definition 2** *For a network with the connectivity graph* $G = (V, E)$, *a cut-set on* $G$ *is defined as a subset* $\mathcal{S} \subseteq V$ *such that* $0 \in \mathcal{S}, K+1 \in \mathcal{S}^c$. *The weight of the cut-set corresponding to* $\mathcal{S}$, *denoted by* $w_G(\mathcal{S})$, *is defined as*

$$w_G(\mathcal{S}) = \sum_{a \in \mathcal{S}, b \in \mathcal{S}^c, (a,b) \in E} N_a N_b. \tag{60}$$

**Theorem 6** *Consider a full-duplex multi-antenna multi-relay network with the graph* $G = (V, E)$ *where* $G$ *is directed acyclic. Assume each node has* $N$ *antennas. The RS scheme achieves the following DMT*

$$d_{RS}(r) = \min_{\mathcal{S}} w_G(\mathcal{S}) \frac{d_{N \times N}(r)}{N^2}, \tag{61}$$

*where* $d_{N \times N}(r)$ *denotes the DMT of a* $N \times N$ *multi-antenna channel.*

*Proof:* Using the same path sequence as the one in the proof of *Remark 6* in [1] and applying the result of Corollary 1, (61) can be derived. The steps are the same as the steps in the proof of Theorem 3, noting the equivalent point-to-point channel is block lower-triangular and the function $d_{N \times N}(r)$ is convex decreasing. ∎

*Remark 1*- According to (61), a specific RS scheme with fixed path and timing sequences can simultaneously achieve the maximum diversity gain, which is $\min_{\mathcal{S}} w_G(\mathcal{S})$ (refer to [1]), and the maximum multiplexing gain, which is $N$ in a multi-antenna relay network whose underlying graph is directed acyclic. In other words, the RS scheme is *robust* in the sense that it achieves the corner-points of the optimum DMT with no modification of the scheme parameters.

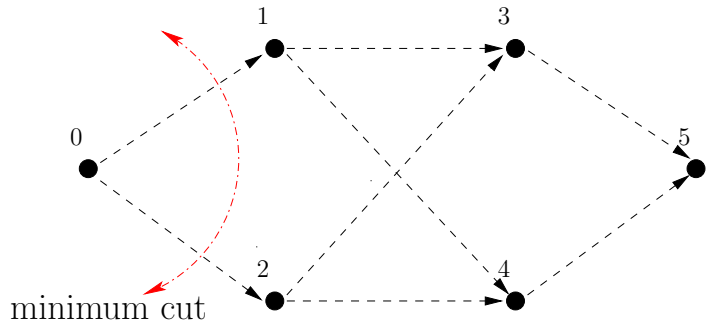

Fig. 5. An example of a multi-antenna directed acyclic network with full-duplex relays, each node equipped with 2 antennas.

Figure 5 shows an example of a directed acyclic network. The relays are operating in the full-duplex mode and each node is equipped with two antennas. Here, the weight of the minimum cut-set depicted in the figure is $8$. Hence, applying the argument of Theorem 6, the RS scheme achieves $d_{RS}(r) = 2d_{2\times 2}(r)$. However, the DMT upper-bound is equal to $d_{ub}(r) = d_{2\times 4}(r)$, which is obtained from the same cut-set. Although the two DMT's are equal in the corner-points, they do not coincide in between.

## IV. Conclusion

In this paper, we derived new DMT results in various setups of multi-antenna relay network using AF relaying. For this purpose, the application of RS scheme proposed in [1] was studied. It was shown that random unitary matrix multiplication at the relay nodes enables the RS scheme to achieve a better DMT in comparison to the traditional AF relaying. First, the multi-antenna full-duplex single-relay two-hop network was studied for which the RS scheme was shown to achieve the optimum DMT. This result was also generalized to the multi-antenna multi-hop full-duplex relay network with one relay in each hop. Next, applying this result, a new achievable DMT was derived for multi-antenna half-duplex parallel relay networks. Interestingly, it turned out that the DMT of the RS scheme is optimum for the multi-antenna half-duplex parallel two-relay setup with no direct link between the relays. Moreover, using random unitary matrix multiplication was shown to improve the DMT of the NAF scheme of [2] in the setup of the widely used multi-antenna single relay channel. Finally, the general full-duplex multi-antenna relay network was studied and a new lower-bound was obtained on DMT using the RS scheme, assuming that the underlying network graph is directed acyclic.

## Appendix I

## Proof of Lemma 1

Assuming a multiplexing gain of $r$, the diversity of the original system can be written as

$$d(r) = \lim_{P \to \infty} -\frac{\log\left(\mathbb{P}\left\{\log\left|\mathbf{I} + \alpha^2 P \mathbf{G}\mathbf{H}\mathbf{H}^H \mathbf{G}^H \left(\mathbf{I} + \alpha^2 \mathbf{G}\mathbf{G}^H\right)^{-1}\right| < r\log(P)\right\}\right)}{\log(P)}. \tag{62}$$

Since $\alpha^2 \mathbf{G}\mathbf{G}^H \succeq \mathbf{0}$, it follows that $\left| \mathbf{I} + \alpha^2 P \mathbf{G}\mathbf{H}\mathbf{H}^H \mathbf{G}^H \left( \mathbf{I} + \alpha^2 \mathbf{G}\mathbf{G}^H \right)^{-1} \right| < \left| \mathbf{I} + \alpha^2 P \mathbf{G}\mathbf{H}\mathbf{H}^H \mathbf{G}^H \right|$. This implies that

$$d(r) \leq d_u(r) \triangleq \lim_{P \to \infty} -\frac{\log \left( \mathbb{P} \left\{ \log \left| \mathbf{I} + \alpha^2 P \mathbf{G}\mathbf{H}\mathbf{H}^H \mathbf{G}^H \right| < r \log(P) \right\} \right)}{\log(P)}, \tag{63}$$

which is the DMT of the system in (2). Moreover, selecting $\alpha = \min \left( 1, \frac{P}{\mathbb{E}\{\|\mathbf{H}\mathbf{x}_t + \mathbf{n}_r\|^2\}} \right)$ (which guarantees that the output power of the relay remains bellow $P$)[12], we have

$$\alpha^2 \mathbf{G}\mathbf{G}^H \preceq \|\mathbf{G}\|^2 \mathbf{I}. \tag{64}$$

Defining the event $\mathscr{C} \equiv \left\{ \|\mathbf{G}\|^2 > c \log(P) \right\}$, and noting that $\|\mathbf{G}\|^2$ is a Chi-square random variable with $2pn$ degrees of freedom, we have

$$\begin{aligned} \mathbb{P}\{\mathscr{C}\} &= \sum_{k=0}^{pn-1} \frac{(c \log(P))^k}{k!} e^{-c \log(P)} \\ &\sim d (\log(P))^{pn-1} P^{-c}, \end{aligned} \tag{65}$$

where $d = \frac{c^k}{k!}$. Defining the outage event of the system in (1) as $\mathcal{O}$, we have

$$\begin{aligned} \mathbb{P}\{\mathcal{O}\} &= \mathbb{P}\{\mathcal{O}|\mathscr{C}\}\mathbb{P}\{\mathscr{C}\} + \mathbb{P}\{\mathcal{O}|\mathscr{C}^c\}\mathbb{P}\{\mathscr{C}^c\} \\ &\leq \mathbb{P}\{\mathscr{C}\} + \mathbb{P}\{\mathcal{O}|\mathscr{C}^c\}\mathbb{P}\{\mathscr{C}^c\}. \end{aligned} \tag{66}$$

Conditioned on $\mathscr{C}^c$, the probability of the outage event can be upper-bounded as

$$\mathbb{P}\{\mathcal{O}|\mathscr{C}^c\} \leq \mathbb{P} \left\{ \log \left| \mathbf{I} + \alpha^2 \frac{P}{c \log P + 1} \mathbf{G}\mathbf{H}\mathbf{H}^H \mathbf{G}^H \right| < r \log(P) \, \middle| \, \mathscr{C}^c \right\}, \tag{67}$$

which is equal to the probability of the outage event of the system in (3), denoted as $\mathcal{O}_l$, conditioned on $\mathscr{C}^c$. In other words, $\mathbb{P}\{\mathcal{O}|\mathscr{C}^c\} \leq \mathbb{P}\{\mathcal{O}_l|\mathscr{C}^c\}$. Substituting in (66) yields

$$\begin{aligned} \mathbb{P}\{\mathcal{O}\} &\leq \mathbb{P}\{\mathscr{C}\} + \mathbb{P}\{\mathcal{O}_l|\mathscr{C}^c\}\mathbb{P}\{\mathscr{C}^c\}. \\ &\leq \mathbb{P}\{\mathscr{C}\} + \mathbb{P}\{\mathcal{O}_l\}. \end{aligned} \tag{68}$$

Since the capacity of the two-hop network is equal to the capacity of both the source-relay and the relay-destination links, it follows that the outage event $\mathcal{O}$ includes $\mathcal{O}_{sr}$, the event of outage in the source-relay link, and $\mathcal{O}_{rd}$, the event of outage in the relay-destination link. As a result,

$$\begin{aligned} \mathbb{P}\{\mathcal{O}\} &\geq \max \left\{ \mathbb{P}\{\mathcal{O}_{sr}\}, \mathbb{P}\{\mathcal{O}_{rd}\} \right\} \\ &\overset{(a)}{\geq} \max \left\{ P^{-mp}, P^{-pn} \right\} \\ &\dot{\geq} P^{-pn}. \end{aligned} \tag{69}$$

$(a)$ results from the fact that the outage probability corresponding to the multiplexing gain $r$ is greater than or equal to the outage probability corresponding to the multiplexing gain 0. Setting $c = 2pn - 1$, from (65), it is concluded that

$$\frac{\mathbb{P}\{\mathscr{C}\}}{\mathbb{P}\{\mathcal{O}\}} = \Theta \left( \left( \frac{\log(P)}{P} \right)^{pn-1} \right). \tag{70}$$

---

[12]Note that as we are looking for a lower-bound for the DMT, it is fine to select any arbitrary value for $\alpha$.

From the above equation and (68), it follows that

$$\mathbb{P}\{\mathcal{O}\} \doteq \leq \mathbb{P}\{\mathcal{O}_l\},\tag{71}$$

which incurs that the DMT of the original system is lower-bounded by the DMT of the system in (3). This completes the proof of Lemma 1.

## APPENDIX II

### PROOF OF LEMMA 2

Defining $P' = \frac{P}{c\log(P)+1}$, the DMT of the system in (3) can be written as

$$
\begin{aligned}
d_l(r) & = \lim_{P \to \infty} -\frac{\log\left(\mathbb{P}\left\{\log\left|\mathbf{I} + \alpha^2 P'\mathbf{G}\mathbf{H}\mathbf{H}^H\mathbf{G}^H\right| < r\log(P)\right\}\right)}{\log(P)} \\
& \overset{(a)}{=} \lim_{P' \to \infty} -\frac{\log\left(\mathbb{P}\left\{\log\left|\mathbf{I} + \alpha^2 P'\mathbf{G}\mathbf{H}\mathbf{H}^H\mathbf{G}^H\right| < r'\log(P')\right\}\right)}{\log(P')} \\
& \overset{(b)}{=} d_u(r') \\
& \overset{(c)}{=} d_u(r),
\end{aligned}
\tag{72}
$$

where $r' \triangleq r\frac{\log(P)}{\log(P')}$ and $(a)$ follows from the fact that as $P' = \frac{P}{c\log(P)+1}$, we have $P = P'\left[c\log(P') + O(\log\log(P'))\right]$, which implies that $\lim_{P \to \infty} \frac{\log(P')}{\log(P)} = 1$. In other words, in the first line of the preceding equation, one can substitute $\log(P)$ by $\log(P')$. $(b)$ results from the fact that the second line in the right hand side of the preceding equation is exactly the DMT of the system in (2) at $r'$, which is denoted by $d_u(r')$. Finally, $(c)$ follows from the facts that $\lim_{P \to \infty} \frac{r'}{r} = 1$ and the continuity of the DMT curve which implies that $d_u(r') = d_u(r)$. This completes the proof of Lemma 3.

## APPENDIX III

### PROOF OF LEMMA 3

First, notice that for values of $i > \min\{m, p\}$ or $i > \min\{p, n\}$, $\lambda_i(\mathbf{G})$ or $\lambda_i(\mathbf{H})$ are defined as zero. Hence, the argument of this lemma is obvious in these cases. Now, we prove the argument for $i \leq \min\{m, n, p\}$. According to the Courant-Fischer-Weyl Theorem [23], we have

$$\lambda_i(\mathbf{G}\mathbf{\Theta}\mathbf{H}) = \max_{\mathcal{S}, \dim(\mathcal{S})=i} \min_{\mathbf{x} \in \mathcal{S}} \frac{\|\mathbf{G}\mathbf{\Theta}\mathbf{H}\mathbf{x}\|^2}{\|\mathbf{x}\|^2}.\tag{73}$$

Now, let us define $\mathcal{S}_0 = \langle \mathbf{v}_1(\mathbf{H}), \mathbf{v}_2(\mathbf{H}), \ldots, \mathbf{v}_i(\mathbf{H}) \rangle$ where $\mathbf{v}_j(\mathbf{H})$ denotes the $j$'th column of $\mathbf{V}(\mathbf{H})$ and $\langle \mathbf{a}_1, \mathbf{a}_2, \ldots, \mathbf{a}_k \rangle$ denotes the span of $\mathbf{a}_1, \mathbf{a}_2, \ldots, \mathbf{a}_k$. We have

$$
\begin{aligned}
\lambda_i(\mathbf{G}\boldsymbol{\Theta}\mathbf{H}) &\geq \min_{\mathbf{x} \in \mathcal{S}_0} \frac{\|\mathbf{G}\boldsymbol{\Theta}\mathbf{H}\mathbf{x}\|^2}{\|\mathbf{x}\|^2} \\
&\overset{(a)}{=} \min_{\mathbf{x}' \in \mathbb{C}^i} \frac{\left\|\mathbf{G}\boldsymbol{\Theta}\mathbf{H}\mathbf{V}_{(1,i)}(\mathbf{H})\mathbf{x}'\right\|^2}{\|\mathbf{x}'\|^2} \\
&= \min_{\mathbf{x}' \in \mathbb{C}^i} \frac{\left\|\mathbf{G}\boldsymbol{\Theta}\mathbf{U}(\mathbf{H})\boldsymbol{\Lambda}^{\frac{1}{2}}(\mathbf{H})\mathbf{I}_{n \times i}\mathbf{x}'\right\|^2}{\|\mathbf{x}'\|^2} \\
&= \min_{\mathbf{x}' \in \mathbb{C}^i} \frac{\left\|\mathbf{G}\boldsymbol{\Theta}\mathbf{U}(\mathbf{H})\boldsymbol{\Lambda}^{\frac{1}{2}}_{(1,i)}(\mathbf{H})\mathbf{x}'\right\|^2}{\|\mathbf{x}'\|^2} \\
&= \min_{\substack{\mathbf{x}' \in \mathbb{C}^i \\ \|\mathbf{x}'\|^2 \geq 1}} \left\|\mathbf{G}\boldsymbol{\Theta}\mathbf{U}(\mathbf{H})\boldsymbol{\Lambda}^{\frac{1}{2}}_{(1,i)}(\mathbf{H})\mathbf{x}'\right\|^2 \\
&= \min_{\substack{\mathbf{x}' \in \mathbb{C}^i \\ \|\mathbf{x}'\|^2 \geq 1}} \left\|\mathbf{G}\boldsymbol{\Theta}\mathbf{U}_{(1,i)}(\mathbf{H})\boldsymbol{\Lambda}^{\frac{1}{2}}_i(\mathbf{H})\mathbf{x}'\right\|^2 \\
&\overset{(b)}{\geq} \min_{\substack{\mathbf{y}' \in \mathbb{C}^i \\ \|\mathbf{y}'\|^2 \geq 1}} \lambda_i(\mathbf{H}) \left\|\mathbf{G}\boldsymbol{\Theta}\mathbf{U}_{(1,i)}(\mathbf{H})\mathbf{y}'\right\|^2 \\
&\overset{(c)}{=} \min_{\substack{\mathbf{y}' \in \mathbb{C}^i \\ \|\mathbf{y}'\|^2 \geq 1}} \lambda_i(\mathbf{H}) \left\|\boldsymbol{\Lambda}^{\frac{1}{2}}(\mathbf{G})\mathbf{V}^H(\mathbf{G})\boldsymbol{\Theta}\mathbf{U}_{(1,i)}(\mathbf{H})\mathbf{y}'\right\|^2 \\
&\overset{(d)}{\geq} \min_{\substack{\mathbf{y}' \in \mathbb{C}^i \\ \|\mathbf{y}'\|^2 \geq 1}} \lambda_i(\mathbf{H}) \left\|\boldsymbol{\Lambda}^{\frac{1}{2}}_{(1,i)}(\mathbf{G})\mathbf{V}^H_{(1,i)}(\mathbf{G})\boldsymbol{\Theta}\mathbf{U}_{(1,i)}(\mathbf{H})\mathbf{y}'\right\|^2 \\
&\overset{(e)}{\geq} \min_{\substack{\mathbf{y}' \in \mathbb{C}^i \\ \|\mathbf{y}'\|^2 \geq 1}} \lambda_i(\mathbf{G})\lambda_i(\mathbf{H}) \left\|\mathbf{V}^H_{(1,i)}(\mathbf{G})\boldsymbol{\Theta}\mathbf{U}_{(1,i)}(\mathbf{H})\mathbf{y}'\right\|^2 \\
&\overset{(f)}{=} \lambda_i(\mathbf{G})\lambda_i(\mathbf{H})\lambda_{\min}\left(\mathbf{V}^H_{(1,i)}(\mathbf{G})\boldsymbol{\Theta}\mathbf{U}_{(1,i)}(\mathbf{H})\right),
\end{aligned}
\tag{74}
$$

where $\mathbf{I}_{n \times i}$ denotes the diagonal identity $n \times i$ matrix and $\boldsymbol{\Lambda}^{\frac{1}{2}}_i(\mathbf{H})$ denotes the square submatrix of $\boldsymbol{\Lambda}^{\frac{1}{2}}(\mathbf{H})$ consisting of its first $i$ rows and first $i$ columns. Here, $(a)$ follows from the fact that i) $\mathbf{x} \in \mathcal{S}_0$ is equivalent to $\mathbf{x} = \mathbf{V}_{(1,i)}\mathbf{x}'$ for some $\mathbf{x}' \in \mathbb{C}^i$; and ii) $\|\mathbf{x}\|^2 = \|\mathbf{x}'\|^2$. $(b)$ follows from the fact that for any $\mathbf{x}' \in \mathbb{C}^i, \|\mathbf{x}'\|^2 \geq 1$, defining $\mathbf{y}' = \frac{1}{\sqrt{\lambda_i(\mathbf{H})}}\boldsymbol{\Lambda}^{\frac{1}{2}}_i\mathbf{x}'$, we have $\|\mathbf{y}'\|^2 \geq \|\mathbf{x}'\|^2 \geq 1$. $(c)$ follows from the fact that for any unitary matrix $\mathbf{P}$, we have $\|\mathbf{P}\mathbf{A}\|^2 = \|\mathbf{A}\|^2$. $(d)$ follows from the fact that defining $\mathbf{z}' = \boldsymbol{\Theta}\mathbf{U}_{(1,i)}(\mathbf{H})\mathbf{y}'$, we have

$$
\left\|\boldsymbol{\Lambda}^{\frac{1}{2}}(\mathbf{G})\mathbf{V}^H(\mathbf{G})\mathbf{z}'\right\|^2 = \left\|\boldsymbol{\Lambda}^{\frac{1}{2}}_{(1,i)}(\mathbf{G})\mathbf{V}^H_{(1,i)}(\mathbf{G})\mathbf{z}'\right\|^2 + \left\|\boldsymbol{\Lambda}^{\frac{1}{2}}_{(i+1,m)}(\mathbf{G})\mathbf{V}^H_{(i+1,m)}(\mathbf{G})\mathbf{z}'\right\|^2.
$$

$(e)$ follows from the fact that defining $\mathbf{w}' = \mathbf{V}^H_{(1,i)}(\mathbf{G})\boldsymbol{\Theta}\mathbf{U}_{(1,i)}(\mathbf{H})\mathbf{y}'$, we have $\left\|\boldsymbol{\Lambda}^{\frac{1}{2}}_{(1,i)}\mathbf{w}'\right\|^2 \geq \lambda_i(\mathbf{G})\|\mathbf{w}'\|^2$. Finally, $(f)$ follows from the Courant-Fischer-Weyl Theorem [23]. (74) completes the proof of Lemma.

# Appendix IV

## Proof of Lemma 4

We have

$$
\begin{aligned}
\left| \boldsymbol{\Psi}_{i,l} \boldsymbol{\Psi}_{i,l}^H \right| &= \lambda_{\min}(\boldsymbol{\Psi}_{i,l}) \prod_{j=1}^{i-1} \lambda_i(\boldsymbol{\Psi}_{i,l}) \\
&\overset{(a)}{\leq} \lambda_{\min}(\boldsymbol{\Psi}_{i,l}) \left( \frac{\mathsf{Tr}\left\{ \boldsymbol{\Psi}_{i,l} \boldsymbol{\Psi}_{i,l}^H \right\}}{i-1} \right)^{i-1},
\end{aligned}
\tag{75}
$$

where $(a)$ follows from the *Geometric Inequality* and the fact that $\sum_{j=1}^{i-1} \lambda_i(\boldsymbol{\Psi}_{i,l}) \leq \mathsf{Tr}\left\{ \boldsymbol{\Psi}_{i,l} \boldsymbol{\Psi}_{i,l}^H \right\}$. $\mathsf{Tr}\left\{ \boldsymbol{\Psi}_{i,l} \boldsymbol{\Psi}_{i,l}^H \right\}$ can be upper-bounded as follows:

$$
\begin{aligned}
\mathsf{Tr}\left\{ \boldsymbol{\Psi}_{i,l} \boldsymbol{\Psi}_{i,l}^H \right\} &= \left\| \mathbf{V}_{(1,i)}^H(\mathbf{G}) \boldsymbol{\Theta}_l \mathbf{U}_{(1,i)}(\mathbf{H}) \right\|^2 \\
&\overset{(a)}{\leq} \left\| \mathbf{V}_{(1,i)}(\mathbf{G}) \right\|^2 \left\| \boldsymbol{\Theta}_l \mathbf{U}_{(1,i)}(\mathbf{H}) \right\|^2 \\
&\overset{(b)}{\leq} \left\| \mathbf{V}_{(1,i)}(\mathbf{G}) \right\|^2 \left\| \mathbf{U}_{(1,i)}(\mathbf{H}) \right\|^2 \\
&\overset{(c)}{=} i^2.
\end{aligned}
\tag{76}
$$

In the preceding equation, $(a)$ follows from the the fact that $\|\mathbf{A}\mathbf{B}\|^2 \leq \|\mathbf{A}\|^2 \|\mathbf{B}\|^2$ for any two matrices $\mathbf{A}$ and $\mathbf{B}$. $(b)$ results from the fact that $\|\boldsymbol{\Theta}\mathbf{A}\| = \|\mathbf{A}\|$, for any matrix $\mathbf{A}$ and any unitary matrix $\boldsymbol{\Theta}$. Finally, $(c)$ follows from the fact that as $\mathbf{U}(\mathbf{H})$ and $\mathbf{V}(\mathbf{G})$ are unitary matrices, each of their columns has unit norm. Combining (75) and (76) yields

$$
\left| \boldsymbol{\Psi}_{i,l} \boldsymbol{\Psi}_{i,l}^H \right| \leq \lambda_{\min}(\boldsymbol{\Psi}_{i,l}) \left( \frac{i^2}{i-1} \right)^{i-1},
\tag{77}
$$

which implies that

$$
\lambda_{\min}(\boldsymbol{\Psi}_{i,l}) \geq c \left| \boldsymbol{\Psi}_{i,l} \boldsymbol{\Psi}_{i,l}^H \right|,
\tag{78}
$$

for some constant $c$. Denoting the $j$th column of $\boldsymbol{\Psi}_{i,l}$ as $\boldsymbol{\Psi}_{i,l}^{(j)}$, we write the determinant of $\boldsymbol{\Psi}_{i,l} \boldsymbol{\Psi}_{i,l}^H$ as

$$
\left| \boldsymbol{\Psi}_{i,l} \boldsymbol{\Psi}_{i,l}^H \right| = \prod_{j=1}^{i} \beta_j,
\tag{79}
$$

where $\beta_j$ denotes the square norm of the projection of $\boldsymbol{\Psi}_{i,l}^{(j)}$ over the null-space of the subspace spanned by $\left\{ \boldsymbol{\Psi}_{i,l}^{(s)} \right\}_{s=1}^{j-1}$. Combining (78) and (79) yields

$$
\begin{aligned}
\mathbb{P}\left\{ \lambda_{\min}(\boldsymbol{\Psi}_{i,l}) \leq \varepsilon \right\} &\leq \mathbb{P}\left\{ \prod_{j=1}^{i} \beta_j \leq \kappa \varepsilon \right\} \\
&\leq \mathbb{P}\left\{ \bigcup_{j=1}^{i} \left\{ \beta_j \leq \sqrt{\kappa \varepsilon} \right\} \right\} \\
&\overset{(a)}{\leq} \sum_{j=1}^{i} \mathbb{P}\left\{ \beta_j \leq \sqrt{\kappa \varepsilon} \right\},
\end{aligned}
\tag{80}
$$

where $\kappa = \frac{1}{c}$ and $(a)$ follow from the union bound on the probability. $\boldsymbol{\Psi}_{i,l}^{(s)}$ can be considered as the projection of the $s$th column of the matrix $\mathbf{R}_l \triangleq \boldsymbol{\Theta}_l \mathbf{U}_{(1,i)}(\mathbf{H})$, denoted by $\mathbf{R}_l^{(j)}$, over the $i$-dimensional subspace spanned by $\{\mathbf{V}_t(\mathbf{G})\}_{t=1}^{i}$, which is denoted by $\mathcal{P}$. Now, consider a random unit vector $\mathbf{w}$ in $\mathcal{P}$, which is orthogonal to the first $j-1$ columns of the matrix $\mathbf{R}_l$. Since $\{\mathbf{\Psi}_{i,l}^{(s)}\}_{s=1}^{j-1}$ are the projections of the first $j-1$ columns of $\mathbf{R}_l$ over $\mathcal{P}$, it follows that $\mathbf{w}$ is also orthogonal to $\{\mathbf{\Psi}_{i,l}^{(s)}\}_{s=1}^{j-1}$. In other words, $\mathbf{w}$ belongs to $\mathcal{Q}^{\perp}$, the null-space of the subspace spanned by $\left\{\mathbf{\Psi}_{i,l}^{(s)}\right\}_{s=1}^{j-1}$ [13]. Hence, we have

$$
\begin{aligned}
\beta_j &= \left\| \mathbf{\Psi}_{i,l}^{(j)}{}^H * \mathcal{Q}^{\perp} \right\|^2 \\
&\geq \left| \mathbf{\Psi}_{i,l}^{(j)}{}^H \mathbf{w} \right|^2,
\end{aligned}
\tag{81}
$$

where $\mathbf{a}^H * \mathcal{T}$ denotes the projection of the vector $\mathbf{a}$ over the subspace $\mathcal{T}$. The second line in the preceding equation follows from the fact that the norm of the projection of a vector over a subspace is greater than the norm of the projection of that vector over any arbitrary unit vector in that subspace. Since $\mathbf{\Psi}_{i,l}^{(j)}$ is the projection of $\mathbf{R}_l^{(j)}$ over $\mathcal{P}$, $\mathbf{R}_l^{(j)}$ can be written as

$$
\mathbf{R}_l^{(j)} = \mathbf{\Psi}_{i,l}^{(j)} + \mathbf{\Psi}_{i,l}^{(j)\perp},
\tag{82}
$$

where $\mathbf{\Psi}_{i,l}^{(j)\perp}$ denotes the projection of $\mathbf{R}_l^{(j)}$ over $\mathcal{P}^{\perp}$, the null-space of $\mathcal{P}$. Since $\mathbf{w} \in \mathcal{P}$, it follows that $\mathbf{w}^H \mathbf{\Psi}_{i,l}^{(j)\perp} = 0$. This means that $\mathbf{\Psi}_{i,l}^{(j)}{}^H \mathbf{w} = \mathbf{R}_l^{(j)}{}^H \mathbf{w}$. Combining the above with (81) yields

$$
\beta_j \geq \left| \mathbf{R}_l^{(j)}{}^H \mathbf{w} \right|^2.
\tag{83}
$$

Since $\mathbf{\Theta}_l$ and $\mathbf{U}(\mathbf{H})$ are unitary matrices and $\mathbf{U}(\mathbf{H})$ is isotropically distributed [20], it follows that $\mathbf{\Theta}_l \mathbf{U}(\mathbf{H})$ is an isotropic unitary matrix. This means i) $\left\{\mathbf{R}_l^{(s)}\right\}_{s=1}^{j}$ is an orthonormal set, which implies that $\mathbf{R}_l^{(j)}$ is orthogonal to $\mathbf{R}_l^{(s)}$, $s = 1, \cdots, j-1$, and ii) $\mathbf{R}_l^{(j)}$ is an isotropic unit vector. As a result, $\mathbf{R}_l^{(j)}$ is an isotropic unit vector in the $(p-j+1)$-dimensional subspace perpendicular to $\left\{\mathbf{R}_l^{(s)}\right\}_{s=1}^{j-1}$. Noting that $\mathbf{w}$ is also in this subspace and $\mathbf{R}_l^{(j)}$ and $\mathbf{w}$ are independent of each other, from [24], Lemma 3, the CDF of $Z_j \triangleq \left| \mathbf{R}_l^{(j)}{}^H \mathbf{w} \right|^2$ can be computed as

$$
F_{Z_j}(z) = 1 - (1-z)^{p-j}.
\tag{84}
$$

Combining (80), (83), and (84), it follows that

$$
\begin{aligned}
\mathbb{P}\left\{\lambda_{\min}(\mathbf{\Psi}_{i,l}) \leq \varepsilon\right\} &\leq \sum_{j=1}^{i} F_{Z_j}(\sqrt[i]{\kappa \varepsilon}) \\
&= \sum_{j=1}^{i} \left[ 1 - \left(1 - \sqrt[i]{\kappa \varepsilon}\right)^{p-j} \right] \\
&\overset{(a)}{\leq} \sum_{j=1}^{i} (p-j) \sqrt[i]{\kappa \varepsilon} \\
&= \eta \sqrt[i]{\varepsilon},
\end{aligned}
\tag{85}
$$

where $\eta = i(p - \frac{i+1}{2})\sqrt[i]{\kappa}$. In the above equation, $(a)$ follows from the fact that $(1-x)^n \geq 1 - nx$, $\forall n \geq 0, 0 \leq x \leq 1$. This completes the proof of Lemma 4.

---

[13] Note that as $j \leq i$, $\mathcal{Q}^{\perp}$ has at least dimension of 1.

## Appendix V

## Proof of Lemma 5

We can assume that $f(\mathbf{x})$ is defined such that $\mathbf{x} \leq \mathbf{y} \Rightarrow f(\mathbf{x}) \leq f(\mathbf{y})$; otherwise, we can redefine $f(\mathbf{x})$ as $f(\mathbf{x}) = \inf_{\mathbf{y} \geq \mathbf{x}} f(\mathbf{y})$. Let us define the set $\mathcal{S}(\epsilon)$ for every $\epsilon > 0$ as $\mathcal{S}(\epsilon) \triangleq \left\{ \mathbf{x} \in [0,\infty)^n \,\middle|\, d(\mathbf{x},\mathcal{R}) \leq \sqrt{n}\epsilon, \frac{\mathbf{x}}{\epsilon} \in \mathbb{Z}^n \right\}$ where $d(\mathbf{x},\mathcal{R}) = \inf_{\mathbf{y} \in \mathcal{R}} \|\mathbf{x} - \mathbf{y}\|$. Also, let us define the partial order $\leq$ for two sets $\mathcal{A}, \mathcal{B} \subseteq [0,\infty)^n$ as follows: $\mathcal{A} \leq \mathcal{B}$ iff for every $\mathbf{x} \in \mathcal{B}$ there exists a $\mathbf{y} \in \mathcal{A}$ such that $\mathbf{y} \leq \mathbf{x}$[14]. One can easily verify that $\mathcal{S}(\epsilon) \leq \mathcal{R}$. Now, let us define the set $\mathcal{L}(\epsilon)$ as $\mathcal{L}(\epsilon) \triangleq \{ \mathbf{x} \in \mathcal{S}(\epsilon) \,|\, \nexists \mathbf{y} \in \mathcal{S}(\epsilon), \mathbf{y} < \mathbf{x} \}$, i.e. $\mathcal{L}(\epsilon)$ consists of the minimal members of $\mathcal{S}(\epsilon)$. Notice that the elements of $\mathcal{S}(\epsilon)$ can be mapped to the elements of $\mathbb{Z}_+^n$ such that the partial order $<$ between the real vectors is kept between the corresponding integer vectors[15]. For every subset $\mathcal{A} \subseteq \mathbb{Z}_+^n$ and every $\mathbf{x} \in \mathcal{A}$, there exists a minimal member $\mathbf{y} \in \mathcal{A}$ such that $\mathbf{y} \leq \mathbf{x}$. Accordingly, we have such a property for $\mathcal{S}(\epsilon)$. This means $\mathcal{L}(\epsilon) \leq \mathcal{S}(\epsilon)$. Noticing $\mathcal{S}(\epsilon) \leq \mathcal{R}$, we conclude that $\mathcal{L}(\epsilon) \leq \mathcal{R}$. On the other hand, it is easy to check that $\mathcal{L}(\epsilon)$ is a finite set, i.e. $|\mathcal{L}(\epsilon)| < \infty$. Hence, we have

$$\mathbb{P}\{\mathcal{R}\} \overset{(a)}{\leq} \mathbb{P}\left\{ \bigcup_{\mathbf{x}' \in \mathcal{L}(\epsilon)} (\mathbf{x} \geq \mathbf{x}') \right\} \overset{(b)}{\leq} |\mathcal{L}(\epsilon)| \max_{\mathbf{x}' \in \mathcal{L}(\epsilon)} \mathbb{P}\{\mathbf{x} \geq \mathbf{x}'\} \overset{(b)}{\leq} P^{-\min_{\mathbf{x} \in \mathcal{L}(\epsilon)} f(\mathbf{x})}, \tag{86}$$

where $(a)$ follows from $\mathcal{L}(\epsilon) \leq \mathcal{R}$ and $(b)$ follows from the fact that $|\mathcal{L}(\epsilon)| < \infty$. Now, let us define $h(\epsilon) = \min_{\mathbf{x} \in \mathcal{L}(\epsilon)} f(\mathbf{x})$. For every $\epsilon$, we have $h(\epsilon) < \inf_{\mathbf{x} \in \mathcal{R}} f(\mathbf{x})$ (otherwise, according to (86), the statement of the Lemma is proved). Now, we prove that $\lim_{\epsilon \to 0} h(\epsilon)$ exists and in fact, we have $\lim_{\epsilon \to 0} h(\epsilon) = \inf_{\mathbf{x} \in \mathcal{R}} f(\mathbf{x})$. As $f(\mathbf{x})$ is *uniformly continuous*, there exists a positive function $g(\epsilon)$ ($g(\epsilon) > 0$) such that for all $\mathbf{x}, \mathbf{y}, \|\mathbf{x} - \mathbf{y}\| \leq g(\epsilon)$, we have $|f(\mathbf{x}) - f(\mathbf{y})| \leq \epsilon$. Consider any positive constant $\delta > 0$ and any $\epsilon \leq \frac{g(\delta)}{\sqrt{n}}$. According to the definition, for any $\mathbf{x} \in \mathcal{L}(\epsilon)$, there exists a $\mathbf{y} \in \mathcal{R}$ such that $\|\mathbf{x} - \mathbf{y}\| \leq \sqrt{n}\epsilon \leq g(\delta)$. Accordingly, $|f(\mathbf{x}) - f(\mathbf{y})| \leq \delta$. Hence, we have $h(\epsilon) = \min_{\mathbf{x} \in \mathcal{L}(\epsilon)} f(\mathbf{x}) \geq \inf_{\mathbf{x} \in \mathcal{R}} f(\mathbf{x}) - \delta$. On the other hand, we know $h(\epsilon) \leq \inf_{\mathbf{x} \in \mathcal{R}} f(\mathbf{x})$. This proves $\lim_{\epsilon \to 0} h(\epsilon) = \inf_{\mathbf{x} \in \mathcal{R}} f(\mathbf{x})$. Noticing $\lim_{\epsilon \to 0} h(\epsilon) = \inf_{\mathbf{x} \in \mathcal{R}} f(\mathbf{x})$ and applying (86) proves the Lemma.

[14] It should be noted that $\leq$ has the properties that i) $\mathcal{A} \leq \mathcal{B}$ and $\mathcal{B} \leq \mathcal{C}$, results in $\mathcal{A} \leq \mathcal{C}$; and ii) $\mathcal{A} \leq \mathcal{A}$.

[15] Such a mapping could be as follows: $m(\mathbf{x}) = \frac{\mathbf{x}}{\epsilon}$.